\documentclass[journal,draftcls, 11pt,onecolumn,letterpaper]{IEEEtran}

\makeatletter
\let\IEEEproof\proof
\let\IEEEendproof\endproof
\let\proof\@undefined
\let\endproof\@undefined
\makeatother

\DeclareMathSizes{11}{11}{7}{8}

\usepackage{amsmath,amsthm,amssymb,euscript,amscd,amsbsy,psfrag,cite,color}
\usepackage{graphicx}
\usepackage{graphics}
\usepackage{lscape}
\usepackage{subfigure}
\def\baselinestretch{1.13}

\newtheorem{lemma}{Lemma}
\newtheorem{theorem}{Theorem}

\newtheorem{definition}{Definition}
\newtheorem{remark}{Remark \textrm}

\newcommand{\pr}{\textnormal{Pr}}
\newcommand{\E}{\mathbf{E}}

\newcommand{\diag}{\textnormal{diag}}

\newcommand{\bfA}{\pmb{A}}

\newcommand{\bfx}{\pmb{x}}
\newcommand{\bfb}{\pmb{b}}
\newcommand{\bfc}{\pmb{c}}

\newcommand{\bfpi}{\pmb{\pi}}

\newcommand{\bara}{\bar{a}}

\newcommand{\calF}{\mathcal{F}}
\newcommand{\calP}{\mathcal{P}}

\newcommand{\hatx}{\hat{x}}
\newcommand{\hatW}{\hat{W}}

\newcommand{\hbfx}{\hat{\bfx}}

\newcommand{\ep}{\epsilon}
\newcommand{\disp}{\displaystyle}

\newcommand{\myif}{~  \mbox{if} ~ }

\newcommand{\eqa}{\stackrel{\textnormal{(a)}}{=}}
\newcommand{\eqb}{\stackrel{\textnormal{(b)}}{=}}
\newcommand{\eqc}{\stackrel{\textnormal{(c)}}{=}}
\newcommand{\eqd}{\stackrel{\textnormal{(d)}}{=}}
\newcommand{\ineqa}{\stackrel{\textnormal{(a)}}{\leq}}
\newcommand{\ineqb}{\stackrel{\textnormal{(b)}}{\leq}}
\newcommand{\ineqc}{\stackrel{\textnormal{(c)}}{\leq}}

\newcommand{\arrowP}{\stackrel{\textnormal{P}}{\rightarrow}}

\newcommand{\ba}{\begin{array}}
\newcommand{\ea}{\end{array}}
\newcommand{\ee}{\end{equation}}
\newcommand{\be}{\begin{equation}}
\newcommand{\bt}{\begin{tabular}}
\newcommand{\et}{\end{tabular}}
\newcommand{\beas}{\begin{eqnarray*}}
\newcommand{\eeas}{\end{eqnarray*}}
\newcommand{\bea}{\begin{eqnarray}}
\newcommand{\eea}{\end{eqnarray}}

\renewcommand{\QED}{\QEDopen}

\let \proof \IEEEproof
\let \endproof \IEEEendproof
\begin{document}

\title{\LARGE \bf
Capacity-achieving Feedback Scheme for Gaussian Finite-State Markov Channels with
Channel State Information }

\author{Jialing Liu, Nicola Elia, and Sekhar Tatikonda \thanks{This research was supported by
the National Science Foundation under Grant ECS-0093950.  This
paper was presented in part at the 2004 American Control
Conference (ACC) and the 2004 IEEE International Symposium on
Information Theory (ISIT). }
\thanks{J. Liu was with the Department of Electrical
and Computer Engineering, Iowa State University, Ames, IA 50011
USA.  He is now with Huawei, Rolling Meadows, IL 60008 USA (e-mail:jialing.liu@huawei.com). }
\thanks{N. Elia is with the Department of Electrical
and Computer Engineering, Iowa State University, Ames, IA 50011
USA (e-mail: nelia@iastate.edu). }
\thanks{S. Tatikonda is with the Department of Electrical Engineering, Yale
University, New Haven, CT 06520 USA (e-mail:
sekhar.tatikonda@yale.edu).}
}

\maketitle

\begin{abstract}
In this paper, we propose capacity-achieving communication schemes
for Gaussian finite-state Markov channels (FSMCs) subject to an average channel input power constraint, under the assumption that the transmitters can have access to delayed noiseless output feedback as well as instantaneous or delayed channel state information (CSI).  We show that the proposed schemes reveals connections between feedback
communication and feedback control.

\end{abstract}


\section{Introduction} \label{sec:intro}

There have been many achievements in the study of time-varying fading channels,
in which the fading gains (referred to as channel states) are often modeled as stochastic processes such as i.i.d. processes or Markov
processes; see
\cite{gold:csi97,shamai_csi99,vis99,kavcic_it03,sahai_ge05, gold:fsmcnocsi06,sahai:main,periodfb07,gold07csi,tati:capI}, to list only a few.
In \cite{gold:csi97}, the capacity and optimal code were obtained for a time-varying fading channel with
instantaneous channel state information (CSI) at both the transmitter and receiver, or at the receiver only.  In \cite{shamai_csi99}, the capacities of several time-varying fading channels under various CSI assumptions (imprecise CSI, delayed CSI, etc.) were investigated. In \cite{vis99}, the capacity was characterized for a finite-state Markov channel (FSMC)
with CSI delayed at the transmitter side  (DTCSI) and instantaneous at
the receiver side.
In \cite{periodfb07}, an FSMC with periodic transmitter-side CSI was studied.
  In \cite{gold07csi}, the capacity problems for several classes of time-varying fading channels (block-memoryless, asymptotically block-memoryless, etc.) under causal CSI assumption (perfect or imperfect) were addressed. For time-varying fading channels exhibiting inter-symbol interference (ISI), see e.g. \cite{kavcic_it03,gold:fsmcnocsi06,gold07csi}. For time-varying fading channels with output feedback, see e.g. \cite{sahai_ge05,sahai:main,tati:capI}.

In this paper, we present capacity-achieving communication
schemes for certain time-varying fading channels with delayed noiseless output feedback, subject to an \textit{average} channel input power
constraint.  In particular, the forward link of the channel, namely the link from the transmitter to the receiver, experiences time-varying fading and additive white Gaussian noise (AWGN) but not ISI.  The fading gains, or the channel states, form either an i.i.d. process or a finite-state Markov chain, and are known to the receiver without delay (or effectively, before the receiver processes the block of outputs) and to the transmitter with or without delay.  The reverse link, also known as the feedback channel, enables the transmitter to access exact channel outputs with delay.

The proposed communication schemes over channels with time-varying fading and output feedback generalize, first, the Schalkwijk-Kailath scheme (SK scheme) over channels without time-varying fading but with output feedback \cite{kailath1,kailath2}, and second, the optimal communication schemes over channels with time-varying fading but without output feedback \cite{gold:csi97,vis99}.  In essence, the proposed
communication system for an FSMC consists of a set of
decoupled SK-type subsystems running in parallel, and the subsystems are
multiplexed to share the forward link and reverse link according to the forward-link channel state evolution.  When the channel state process is i.i.d., however, a simplified adaptive scheme without multiplexing can be used to achieve the capacity.

This paper also reveals tight connections between the feedback communication problem over an FSMC and a related feedback stabilization problem over a Markov Jump Linear System (MJLS) that has the same channel in the loop.  We show that, if the MJLS, unstable in the open loop, is stabilized in the closed loop, then its corresponding communication system can achieve a communication rate arbitrarily close to the so called open-loop growth rate, which is a measure of how unstable the MJLS is in the open loop (see Section \ref{sub:control} for details).
Moreover, the transmission power in the communication system can be determined from the MJLS by solving an optimal control problem called the \emph{cheap control}.  Therefore, the optimality in the communication problem, namely the optimal rate versus power relation, can be completely characterized by analyzing the associated control problem, and we show that this leads to a control-oriented approach that may be employed to facilitate the development of capacity-achieving communication schemes.

\textit{Organization:}    Section \ref{sec:model} introduces the
channel models and capacity concepts. In Section \ref{sec:nofading} we review an SK-type system which achieves the feedback capacity of an (a unit-gain) AWGN channel, followed by the optimal scheme for a constant-gain channel with AWGN.  We then present in Section \ref{sec:d=0} the optimal scheme for channels with instantaneous transmitter-side CSI (TCSI).  In Section \ref{sec:iid} we study Gaussian i.i.d fading channels with DTCSI, and in Section \ref{sec:fsmc}, the Gaussian FSMC with DTCSI.  After discussing connections to feedback control problems in Section \ref{sec:control}, we present a numerical example in Section
\ref{sec:eg}.

\textit{Notations:} We represent time indices by subscripts, such as $A_n$; to conform
with the convention in dynamical systems, the time index starts from 0.  We denote by
$A_n^m$ the sequence $\{A_n,A_{n+1}, \cdots, A_m\}$. 
We use boldface letter $\bfx$ for a vector, and $x^{(i)}$ for
the $i$th element of the vector $\bfx$. Note that $A_n^m$ is a sequence, $(A_n)^m$ is the
$m$th power of $A_n$, $\bfA_n$ is a vector with the time index $n$, and $A_n^{(m)}$ is the $m$th
element of the vector $\bfA_n$. We use $a[1], a[2],\cdots$ to represent a collection of
fixed numbers. We denote ``defined to be" as ``$:=$".  The notation $\lfloor x \rfloor$ denotes the largest integer no greater than $x$.  The notation $\arrowP$ specifies convergence in probability.

\section{Channel Models and Capacities}  \label{sec:model}

In this section, we first describe the forward-link and reverse-link channel models, followed by the discussion of CSI assumptions and interconnected channels.   We then present capacity definitions and the capacity theorem.

\subsection{Channel models} \label{sub:models}

\textbf{The forward-link model $F$ and specializations}

\begin{figure}[h!]
\begin{center}
\subfigure[]{\scalebox{.65}{\includegraphics{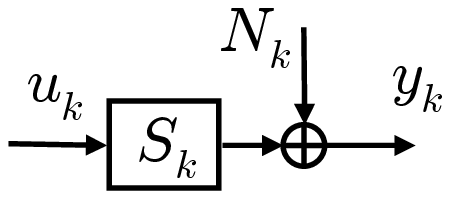}}} \hspace{8pt}
\subfigure[]{\scalebox{.5}{\includegraphics{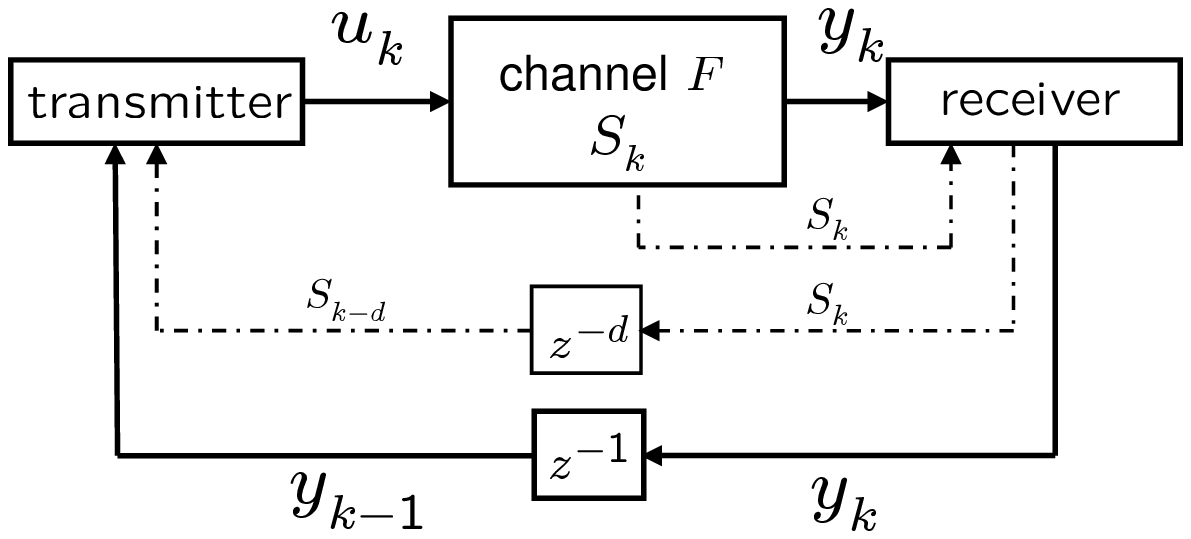}}} \caption{(a) The forward-link channel $F$. (b) The interconnected channel $\calF$ (cf. Definition \ref{def:channels}; it is $\calF_{TCSI}$ if $d=0$ or $\calF_{DTCSI}$ if $d=1$).  } \label{fig:channel}
\end{center}
\end{figure}

The general forward-link channel is depicted in Fig. \ref{fig:channel} (a).  At time $k$ it is
described as
\begin{equation} {F} : \;\;
y_{k} = S_k u_{k} + N_k, \;\textnormal{ for }
k=0,1,2,\cdots,\label{mod:afsmc} \end{equation}
where $u_{k}$ is the channel input, $S_k$ is the channel gain (also known as the channel state),
$N_k$ is the noise, and $y_{k}$ is the output. These
variables are real-valued.  The noise $\{N_k\}$ is independent
Gaussian with zero mean and a unit variance.
We assume $\pr (S_k|S_{k-1},u_0^{k-1},y_0^{k-1})=\pr (S_k|S_{k-1})$, which implies $\pr (y_k|S_{k},u_0^{k})=\pr (y_k|S_{k},u_k)$, i.e., the channel has no ISI.  Furthermore, we assume that $\{S_k\}$ forms a stationary, irreducible, aperiodic, finite-state homogeneous Markov chain and hence is ergodic, with one-step
transition probability
\begin{equation} p_{ij} := \pr (S_k=s[j]|S_{k-1} = s[i]),  \;\textnormal{ for }
k=1,2,\cdots, \label{mctransprob} \end{equation}
where $i,j=1,2,\cdots,m$; $m$ is the number of possible channel state values; and $s[i]$ is a fixed number for each $i$ with $s[i] \neq s[j]$ if $i \neq j$. In this paper $s[i]$ denotes one
of the $m$ channel states, and it also represents the
associated channel gain if the channel is in that state.  We summarize the assumptions in the following definition:

\begin{definition} \label{def:forward}
The forward-link channel $F$ is an ergodic FSMC corrupted by AWGN according to (\ref{mod:afsmc}) and the channel state $S_k$ evolves according to (\ref{mctransprob}).
\end{definition}

Define the one-step transition matrix as $P:=((p_{ij}))$ for the Markov chain. By
ergodicity, the stationary distribution $ \bfpi :=
[\pi[1],\pi[2],\cdots,\pi[m]] $ exists and is the normalized \emph{positive} solution to $\bfpi=\bfpi P$.

Additional assumptions on the channel state $S_k$ may be adopted to represent some more specific, widely used forward-link channel models:

\begin{definition} \label{def:forwardmore}
i) A channel $F$ is called a Gaussian i.i.d fading channel, denoted $F_{I}$, if $p_{ij}=p_{lj}=\pi[j]$ for any $i,j,l$.
ii) A channel $F$ is called a constant-gain AWGN channel, denoted $F_C$, if $m=1$.
iii) A channel $F$ is called an AWGN channel with a unit gain, denoted $F_A$, if $m=1$ and $s[1]=1$.
\end{definition}

The channels $F$, $F_I$, $F_C$, and $F_A$ form a nested relation as the former ones encompass the latter ones.

The channel $F$ may be used to model the following cases and their
generalizations. For one, a continuous-alphabet channel subject to random erasures
(i.e. discrete channel states) and AWGN, in which
the erasures may exhibit certain time
correlation (e.g. forming a two-state Markov chain) as the causes of erasures may be time-correlated. For another, a continuous-alphabet channel subject
to bursty noises with different noise variances, in which the occurrence
of bursty noises forms a finite-state Markov chain.  The well-known
Gilbert-Elliot channel with AWGN falls into this category.    Note that continuous-alphabet channels are widely studied in the literature, especially when output feedback is used.
Note also that the discreteness of the channel states may arise from quantizing continuous channel states (cf. \cite{vis99} and therein references), though the impact of quantization may need further investigation when one applies the coding strategies developed for the induced FSMCs to the original continuous-state channels.

\textbf{The reverse-link model $F_R$}

We denote the reverse link with noiseless, one-step-delayed \emph{output feedback} as $F_R$. That is, the channel input $u_k$ can depend on $y_0^{k-1}$ but not $y_k$.  The noiseless assumption, despite of being practically unrealistic in many systems, is widely adopted and is shown to be useful in establishing conceptually insightful results (cf. e.g. \cite{elia_c5,sahai:main,tati:capI}).  It may also shed light on how the unsolved problem of achieving channel capacity with noisy feedback can be approached.

\textbf{CSI assumptions}

Exact CSI is assumed throughout the paper.  This is not quite realistic but has been shown useful in simplifying the analysis and gaining understandings of the problems under study (see e.g.
\cite{gold:csi97,vis99}). The receiver can access CSI with no delay or effectively, before the receiver processes the current block of channel outputs.  The transmitter can access CSI with no delay (i.e. $d=0$, or TCSI) or with one-step delay (i.e. $d=1$, or DTCSI).  Note that TCSI may be obtained effectively using sounding in a Time Division Duplexing (TDD) system, whereas DTCSI may be obtained by sending the instantaneous receiver-side CSI to the transmitter via the reverse link with one-step delay.  Though it may be feasible that the transmitter has access to instantaneous CSI, it is not feasible that the transmitter has access to instantaneous \emph{output feedback} which would then violate strict causality and lead to an algebraic loop.  \emph{Unless otherwise specified, the term ``feedback'' means output feedback.}

\textbf{Interconnected channels $\mathcal{F}_{TCSI}$, $\mathcal{F}_{DTCSI}$, and specializations}

Combining the above forward-link and reverse-link models with appropriate CSI assumptions, we identify several interconnected channels as shown below, which are generically referred to as $\calF$.

\begin{definition} \label{def:channels}
i) Let $\mathcal{F}_{TCSI}$ be the interconnected channel with the forward link $F$, reverse link $F_R$, and TCSI.
ii) Let $\mathcal{F}_C$ be the interconnected channel $\mathcal{F}_{TCSI}$ with the forward link $F_C$.
iii) Let $\mathcal{F}_A$ be the interconnected channel $\mathcal{F}_{TCSI}$ with the forward link $F_A$.
iv) Let $\mathcal{F}_{DTCSI}$ be the interconnected channel with the forward link $F$,  reverse link $F_R$, and DTCSI.
v) Let $\mathcal{F}_{I,DTCSI}$ be the interconnected channel $\mathcal{F}_{DTCSI}$ with the forward link $F_I$.

\end{definition}

We illustrate these channels in Fig. \ref{fig:channel} (b), in which instantaneous receiver-side perfect CSI is always assumed.

\subsection{Channel capacities} \label{subsec:cap}

\textbf{The operational capacity}

\begin{definition} \label{def:code}

Consider the channel $\calF$. An $(M_K,K+1)$ code with the time span $0,1,\cdots,K$ and  power budget $\calP$ consists of the following:

i) A set of $M_K$ equally likely messages $w_K:=\{w[1], \cdots, w[M_K]\}$ known to both the transmitter and receiver;

ii) An encoding function generating the channel input at the transmitter side as $u_k:=u_k(W_K,y_0^{k-1},S_0^{k-d})$, where $k=0,\cdots,K$, $d=1$ for DTCSI and $d=0$ for TCSI, and $W_K \in w_K$ is the selected message known to the transmitter but not the receiver, subject to the following average transmission power constraint
\begin{equation} \frac{1}{K+1} \sum_{k=0}^K \E (u_k)^2 \leq \calP ; \label{code_pc} \end{equation}

iii) A decoding function generating the decoded message at the receiver side as $\hatW_K:=\hatW_K(y_0^K,S_0^K)$.

The rate of the $(M_K,K+1)$ code is
\begin{equation} R_K:=\frac{1}{K+1} \log M_K , \end{equation}
and the probability of error of the code is $PE_K:=\Pr(\hatW_K \neq W_K)$.
\end{definition}

\begin{definition} \label{def:achievable_rate}
A rate $R$ is said to be achievable with the power budget $\calP$ for a channel if there exists a sequence of $(M_K,K+1)$ codes satisfying the power constraint (\ref{code_pc}) such that $\liminf_{K \rightarrow \infty} R_K \geq R$ and $\lim_{K \rightarrow \infty} PE_K = 0$.
\end{definition}

\begin{definition} \label{def:o_cap}
The operational capacity $C^o(\calP)$ for the channel $\calF$ is the supremum of all achievable rates with the power budget $\calP$.
\end{definition}

\textbf{The information capacity}

The ``information'' channel capacity is defined below as initially characterized in \cite{vis99}.  It is a ``single-letter'' expression, namely it is in terms of the mutual information between one channel input $u$ and one channel output $y$ related as $y=S u+N$ and $u$ depends on $S_{-d}$, where $S_{-d}$ has distribution $\bfpi$ and it transitions to $S$ in $d$ steps with the one-step transition matrix being $P$.

\begin{definition} \label{def:I_cap}
The information channel capacity $C(\calP)$ for the channel $\calF$  is
\begin{equation} \ba{lll} \disp C(\calP) &:=& \disp \max_{\Pr(u|S_{-d})}  \E _{S_{-d} \sim \bfpi,S } I(u;y | S_{-d},S ) ,\ea \end{equation}
where $\Pr(u|S_{-d})$ is any input distribution subject
to the \textit{average} transmission power constraint
\begin{equation}  \E u^2 \leq \mathcal{P} . \label{inf_cap_pc}\end{equation}
\end{definition}


The information channel capacity can be more explicitly computed as
\begin{equation} \ba{lll} \disp C(\calP) &
=& \disp   \max_{\gamma(\cdot):\sum _{j=1} ^m  \pi[j] \gamma(s[j]) \leq \mathcal{P}} \frac{1}{2}\E  _{S_{-d} \sim \bfpi,S }  \log \left( 1 + S^2 \gamma(S_{-d}) \right) \\
&=& \disp   \frac{1}{2}\E  _{S_{-d} \sim \bfpi,S }  \log \left( 1 + S^2 \Gamma(S_{-d}) \right) \ea
\ee
where $\gamma(\cdot)$ is a power allocation function that
maps the channel state $S_{-d}$ to the transmission power
$\gamma(S_{-d})$, and $\Gamma (\cdot)$ is the \emph{optimal} power allocation function.  These expressions were first obtained in \cite{vis99} (Lemma 2, with $d=0$ or 1 and with $\sigma_s^2=1$ therein). The function $\Gamma (\cdot)$ is given by the solution of a set of $m$ equations (see Appendix B in~\cite{vis99}) and is assumed given throughout this paper; these
equations, involving only arithmetic operations, can be readily solved numerically, and since
the optimization variables $\gamma(s[i])$,  $i=1,\cdots,m$, are inside a compact
region, a number of numerical approaches, such as branching-and-bound, are available
to improve the search efficiency.

Furthermore, we can derive
\be \ba{rll} \label{C_d0d1}  C(\calP)
=& \disp \left\{ \ba{ll}   \disp \frac{1}{2} \sum _{j=1} ^m  \pi[j]  \log (1+s[j]^2 \Gamma (s[j]))  & \textnormal{ for } d=0 \\
\disp  \frac{1}{2} \sum _{j=1} ^m  \sum _{l=1} ^m \pi[j] p_{jl} \log (1+s[l]^2 \Gamma (s[j]))  & \textnormal{ for } d=1
  \ea
\right. \\
=& \log \tilde{a} ,
 \ea \end{equation}
where
\be \tilde{a} : =  \disp \prod _{j=1}^m \bar{a}[j]  \label{tildea} \ee
and
\begin{equation}  \ba{rll}\bara[j] &:=& \disp a(s[j])^{\pi[j]} \\
 a(s[j]) &: =& \disp \sqrt{1+s[j]^2 \Gamma (s[j])} 
\label{eq:abar_d0} \ea \end{equation}
for the channel $\calF_{TCSI}$ (i.e. $d=0$), and
\be \ba{rll} \bar{a}[j] &: =& \disp \prod _{l=1}^m a(s[j],s[l])^{\pi[j] p_{jl}} \\
a(s[j],s[l]) &: =& \disp \sqrt{1+s[l]^2 \Gamma (s[j])}
\label{eq:abar_d1}  \ea \ee
for the channel $\calF_{DTCSI}$ (i.e. $d=1$).  As a special case, for the channel $\calF_{I,DTCSI}$, it holds that $p_{jl}=\pi[l]$ and $\Gamma(s[j])=\calP$ for all $j$ and $l$ since no information about the future channel state can be inferred from the delayed CSI and hence a uniform power allocation is optimal (which is readily proven using Jensen's Inequality \cite{shamai_csi99}). Then (\ref{eq:abar_d1}) reduces to
\be \ba{lll} \bara[j] &:=& a(s[j])^{\pi[j]} \\
 a(s[j]) &: =& \sqrt{1+s[j]^2 \calP}  . \ea \label{eq:abar_iid} \ee

It holds that $\tilde{a}>1$, which follows from the following lemma:

\begin{lemma} \label{lemma:bara}

For any $j=1,\cdots,m$, it holds that $\bar{a}[j] = 1$ if and only if $\Gamma (s[j])=0$.
\end{lemma}
\proof The ``if'' direction is straightforward.  For ``only if'', in the TCSI case, the condition $\bar{a}[j] = 1$ leads to either i) $s[j]=0$ or ii) $\Gamma (s[j])=0$, but i) also implies ii) to ensure optimal power allocation (i.e. no power should be used if the transmitter knows no information can be transmitted).  For the DTCSI case, assume on the contrary $\Gamma (s[j])>0$.   It must then hold that $a(s[j],s[l])^{p_{jl}} =1$ for all $l$, which implies that for each $l$, we either have i) $s[l]=0$ or ii) $p_{jl}=0$.  In other word, there must be a state $s[i]=0$, and for any $l \neq i$, it holds that $p_{jl}=0$.  Hence, we have $p_{ji}=1$, resulting in $\Gamma(s[j])=0$, a contradiction.  Thus $\Gamma (s[j])=0$ follows.   \endproof

\textbf{The channel coding theorem}

\begin{theorem} \label{th:codingTh}
$C^o(\calP) = C(\calP)$.

\end{theorem}

In \cite{vis99}, the capacity theorem for discrete FSMCs was proven, and it was correctly pointed out that the capacity theorem still holds for Gaussian FSMCs. An explicit proof, however, was not included in \cite{vis99} and is presented here.

\proof The converse part is proven in Appendix \ref{app:converse}.  The main idea of the achievability proof is to decompose the channel $\calF$ into a set of parallel channels activated in different time instants, and at each time adapt the coding strategy based on relevant CSI.  The proof may employ random codes \emph{without utilizing output feedback at the transmitter} similar to the proof in Sec. III-B of \cite{vis99}; the detail is skipped for brevity.   Alternatively, we will show in Theorems \ref{th:capd0},  \ref{th:capiid}, and \ref{th:cap} that the explicitly constructed schemes \emph{utilizing output feedback at the transmitter} achieve $C(\calP)$. (Since $C(\calP)$ is achieved whether output feedback is used or not, we see that output feedback does not provide any capacity advantage for the channel $\calF$.) \endproof

We remark that the decomposition of the channel under study into parallel channels cannot be done for channels with ISI such as the FSMC considered in \cite{kavcic_it03} and therefore our analysis and results do not apply to those channels.

\textbf{Comments on the power adaptation}

The power adaptation at the transmitter according to the available CSI and channel correlation has been studied in the literature for channels without output feedback (see e.g. \cite{gold:csi97,shamai_csi99,vis99}).  It has been shown that for channels with TCSI, power adaptation according to the latest TCSI is optimal, independent of whether the channel is i.i.d. or Markov.  For i.i.d. channels with DTCSI, since the DTCSI does not provide any information about the channel state to be experienced, a uniform power allocation is optimal.  For FSMCs with DTCSI, however, power adaptation according to the latest DTCSI is optimal.  These power adaptation strategies will be employed in later sections and we will see that they are still optimal for channels with output feedback.

\section{The optimal schemes for channels $\calF_A$ and $\calF_C$}
\label{sec:nofading}

In this section, we review an optimal
communication scheme over the channel $\calF_A$, which is a minor variation of the ingenious codes initially proposed by Schalkwijk and Kailath (cf. e.g. \cite{kailath1,kailath2,gallager,elia_c5,sahai:main}).  With some further modifications, this SK-type scheme is also capacity-achieving for the channel $\calF_C$.  As we will see in later sections, generalizations of the SK-type scheme can solve the capacity-achieving problems for the general channel $\calF$.

\subsection{The optimal scheme over the channel $\calF_A$} \label{sub:awgnscheme}

\begin{figure}[h!]
\begin{center}
\scalebox{.55}{\includegraphics{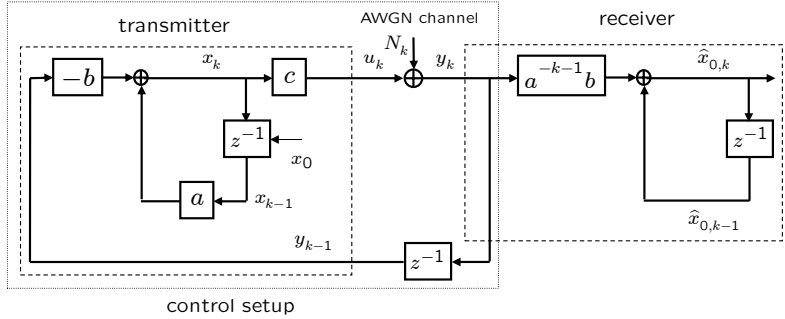}} \caption{ The
communication system for the channel $\calF_A$. The system inside the dotted box represents a closed-loop
control system and will be discussed in Sec. \ref{sec:control}.
 }
\label{fig:awgn}
\end{center}
\end{figure}

Fig. \ref{fig:awgn} shows the optimal communication system for the channel $\calF_A$.  Fix any $\ep>0$ where $\ep$ is an arbitrarily small slack from the capacity.   In what follows we will construct an $(M_K,K+1)$ code with a power budget $\calP>0$ (see Definition \ref{def:code}), where
\begin{equation} M_K:=\lfloor a^{(K+1)(1-\ep)} \rfloor \label{choice:skMk}   \end{equation}
and
\begin{equation} a:=\sqrt{1+\calP}; \label{choice:sk}\end{equation}
namely, the message set $w_K$ has $M_K$ messages.  Define a codebook $v_K$ as $v_K := \{ v[i] | v[i]:=-\sqrt{\calP}+ 2(i-1)D_K,i=1,\cdots,M_K \}$ where
\begin{equation}
 D_K := \disp \frac{\sqrt{\calP}}{ M_K -1} .\label{choice:skDk}  \end{equation}
Therefore, there is a 1-to-1 correspondence between $w_k$ and $v_k$, and the codewords are points in the interval $[-\sqrt{\calP},\sqrt{\calP}]$ with a uniform spacing $2D_K$ between any two neighbors.  Reveal $w_K$ and $v_K$ to both the transmitter and receiver \textit{a priori}.   

To encode, suppose that ones wishes to convey a randomly selected message $W_K \in w_K$, and the corresponding codeword is $V_K \in v_K$. Let
\be b:=\frac{\calP}{a}, \quad c:=1. \label{choice:skbc} \ee
Generate a sequence of channel inputs $u_0^K$ recursively according to:
\begin{equation} \label{dyn:awgncommu:enc}
\begin{array}{lll}
x_k &=& a x_{k-1}-b y_{k-1} \\
u_{k} &=& c x_{k}  ,
\end{array}
\end{equation}
with $y_{-1}:=0$ and $x_{-1}:=V_K/a$, i.e., $x_0=V_K$, implying that the initial condition (at time $k=0$) of the transmitter is the selected codeword $V_K$.    We call $x_k$ the {system state}.
To decode, first based on the channel outputs $y_0^K$ generate the \emph{decoder estimate} $\hatx_{0,k}$ according to:
\begin{equation} \label{dyn:awgncommu:dec}
\hatx_{0,k} = \hatx_{0,k-1} + a^{-k-1}b y_k
\end{equation}
with $\hatx_{0,-1}:=0$.  One can then decode by mapping $\hatx_{0,K}$ into
the closest codeword $\hat{V}_K$ (closest in the sense of the Euclidean distance) to obtain the decoded message $\hat{W}_K$.

The asymptotic rate of the sequence of $(M_K,K+1)$ code is
\begin{equation} \ba{lll} R &=& \disp \lim _{K \rightarrow \infty} \frac{1} {K+1}\log  M_K  \\
&=& \disp \lim _{K \rightarrow \infty} \frac{1} {K+1}(\log  a^{(K+1)(1-\ep)} - \log \xi_K)  \\
&=& (1-\ep)\log a
, \ea \label{awgn_rate}\end{equation}
where we have defined $\xi_K:=a^{(K+1)(1-\ep)}/ M_K $ and used the fact that $\xi_K  \in [1,2)$ for all $K$ since $a>1$.

We note that the formulation of the coding scheme is a variation of the original formulation of the SK scheme.  See Appendix \ref{app:variations} for more discussions.

\subsection{Proof of the optimality of the coding scheme for $\calF_A$ }

We show that this SK-type scheme achieves the rate $R$ in (\ref{awgn_rate}) for any $\epsilon >0$.

\textbf{The closed-loop equation and end-to-end equation}

Using (\ref{choice:sk}) and (\ref{dyn:awgncommu:enc}), we obtain the following equation referred to as the closed-loop equation:
\begin{equation} x_k = a^{-1} x_{k-1} - b N_{k-1}. \label{cl_awgn}\end{equation}
One can then show that the equation from the codeword $V_K$ (or equivalently $x_0$) to the receiver estimate $\hatx_{0,k}$, which we may call as the \emph{end-to-end equation}, is
\begin{equation} \hatx_{0,k} = (1 - a^{-2k-2}) x_0 +  a^{-2k-2} \left( \sum_{t=0}^k a^{t+1} b N_t \right). \label{eq:hbfx0}
  \end{equation}
To see this, recursively apply (\ref{dyn:awgncommu:enc}) and (\ref{cl_awgn}) to obtain
\be \ba{lll} x_k&=& \disp a^k x_0 - a^k \sum _{t=0} ^ {k-1}  a^{-t-1} b y_t \\
&=&  \disp a^{-k} x_0 - a^{-k} \sum _{t=0} ^ {k-1} a^{t+1} b  N_t .  \label{eq:states1} \ea \ee
Then (\ref{eq:hbfx0}) follows from
\be  \hatx_{0,k-1} = \disp \sum _{t=0} ^{k-1} a^{-t-1} b y_t
= \disp x_0-a^{-k} x_k  
.   \ee

\textbf{The Average input power}

By (\ref{eq:states1}), it holds that
\begin{equation} \ba{lll} \E (x_k)^2 &=& \disp a^{-2k}\E (x_0)^2 + \sum_{t=0}^{k-1} a^{-2k+2+2t} b^2  \\
&=& a^{-2k}\left(\E (x_0)^2-\calP \right) + \calP \leq \calP,
\ea  \end{equation}
where the last inequality is due to $|x_0| \leq \sqrt{\calP}$. Since $u_k=x_k$,
the time-average of the input power $ \E (u_0^K{}' u_0^K )/(K+1) $ does not exceed the budget $\calP$.

\textbf{The probability of error}

The end-to-end equation (\ref{eq:hbfx0}) implies that $\hatx_{0,K}$ is Gaussian conditioned on $x_0$:
\begin{equation} \hatx_{0,K|x_0} \sim \mathcal{N}\left((1 - a^{-2K-2}) x_0, \left( a^{-K-1} \sqrt{ (1 - a^{-2K-2}) \calP } \right) ^2 \right) . \label{dist:sk} \end{equation}
Denote the mean as $\mu_K$ and variance $(\sigma_K)^2$.
Therefore, it holds that
\begin{equation} \ba{lll} \disp PE_{K|x_0} &\leq& \disp\Pr (\hatx_{0,K} \geq x_0 + D_K ) + \Pr (\hatx_{0,K} \leq x_0 - D_K) \\
&=& \disp Q\left( \frac{ D_K +x_0 - \mu_K }{\sigma_K}\right) + Q\left( \frac{D_K -x_0 + \mu_K }{\sigma_K}\right) \\
&=& \disp  Q\left(  \frac{1}{\sqrt{ 1 - a^{-2K-2}}} \left( \frac{a^{K+1}}{ \lfloor a^{(K+1)(1-\ep)} \rfloor -1 } +  a^{-K-1} \frac{x_0}{\sqrt{\calP}}  \right) \right)  \\
 && \disp + Q\left(  \frac{1}{\sqrt{ 1 - a^{-2K-2}}} \left( \frac{a^{K+1}}{ \lfloor a^{(K+1)(1-\ep)} \rfloor -1 } -  a^{-K-1} \frac{x_0}{\sqrt{\calP}}  \right) \right) \\
&\leq& \disp 2 Q\left(  \frac{1}{\sqrt{ 1 - a^{-2K-2}}} \left( \frac{a^{K+1}}{ \lfloor a^{(K+1)(1-\ep)} \rfloor -1 } -  a^{-K-1}  \right) \right)  , \ea
\end{equation}
where $Q(\cdot)$ is the Gaussian Q-function.  The first inequality (as opposed to equality) is because when, say, $x_0:=\sqrt{\calP}$, then any noise such that $\hat{x}_{0,K}>x_0$ would not result in a decoding error.  The last inequality is because the Q-function is strictly decreasing and $|x_0|\leq \sqrt{\calP}$.  Since $a>1$, straightforward computation can show that as $K$ tends to infinity, the above upper bound of $PE_{K|x_0}$, which is independent of $x_0$, decreases as $2Q(a^{(K+1)\ep})$ which goes to zero.  This then follows that $PE_K \rightarrow 0$.
Thus, any rate below the capacity is achievable by this scheme.

\begin{remark} \label{rem:unbiased}
We may employ a modified decoding method by mapping $(1-a^{-2K-2})^{-1}\hat{\bfx}_{0,K}$ into the closest codeword to obtain the decoded message, which removes the estimation bias (i.e. the term $-a^{-2K-2}x_0$ in (\ref{dist:sk})) and also leads to reliable communication \cite{gallager}.
\end{remark}

\subsection{The optimal scheme over the channel $\calF_C$}
With some minor modifications, the optimal scheme for the channel $\calF_A$ can achieve any rate below the capacity (proof omitted).  To this aim, one needs to only change parameters $a$ and $b$ in (\ref{choice:sk}) and (\ref{choice:skbc}) to
\begin{equation} a:=\sqrt{1+s^2\calP}, \quad b:=\frac{\calP s}{a},  \label{choice:FC}\end{equation}
where $s$ is the constant gain of the forward link.  Equation (\ref{choice:FC}) indicates that the transmitter and receiver parameters need to appropriately reflect the channel gain in order to achieve the capacity.  It is then expected that the transmitter and receiver need to adapt to the time-varying CSI if a time-varying channel is considered, as we will see in later sections.

\section{The optimal scheme for the channel $\calF_{TCSI}$} \label{sec:d=0}

In this section, we present the optimal feedback communication scheme for the channel $\calF_{TCSI}$.  The proposed system is a \emph{multiplexed adaptive system with power adaptation}.  The main idea behind the scheme is to build parallel subsystems and multiplex among them according to the CSI such that each subsystem sees only a constant channel state, similar to the case without output feedback.  More specifically, one can decompose $\calF_{TCSI}$ into a set of $m$ parallel sub-channels.  Then the sub-channel associated with the channel state $s[l]$ may be viewed as a constant-gain channel, over which one can construct an SK-type system referred to as the subsystem $\Sigma_l$.  At time $k$ the subsystem $\Sigma_l$ transmits over the forward link if and only if $S_k=s[l]$, in the meantime it sends the channel output via the reverse link.  The output feedback will reach the transmitter at time $k+1$, and will be fed to $\Sigma_l$ at time $k+1$ (i.e. at time $k+1$ the transmitter needs the delayed CSI $S_k$ in order to correctly feed the output feedback to $\Sigma_l$).  
It follows that the subsystem $\Sigma_l$ can achieves its capacity $C_l=\frac{1}{2} \log(1+s[l]^2\Gamma(s[l]))$ if its power budget is $\Gamma(s[l])$.  Then by ergodicity of the channel state process, the $m$ decoupled subsystems, when multiplexed according to the CSI, can achieve the capacity $C=\log \tilde{a}$.

Since the correlation between the channel states does not provide any additional information under the TCSI assumption, the result in this section is applicable to $\calF_{TCSI}$ with either i.i.d or Markov channel state process in its forward link.


\subsection{The proposed communication system} \label{subsec:schemed=0}

\begin{figure}[h!]
\begin{center}
{\scalebox{.55}{\includegraphics{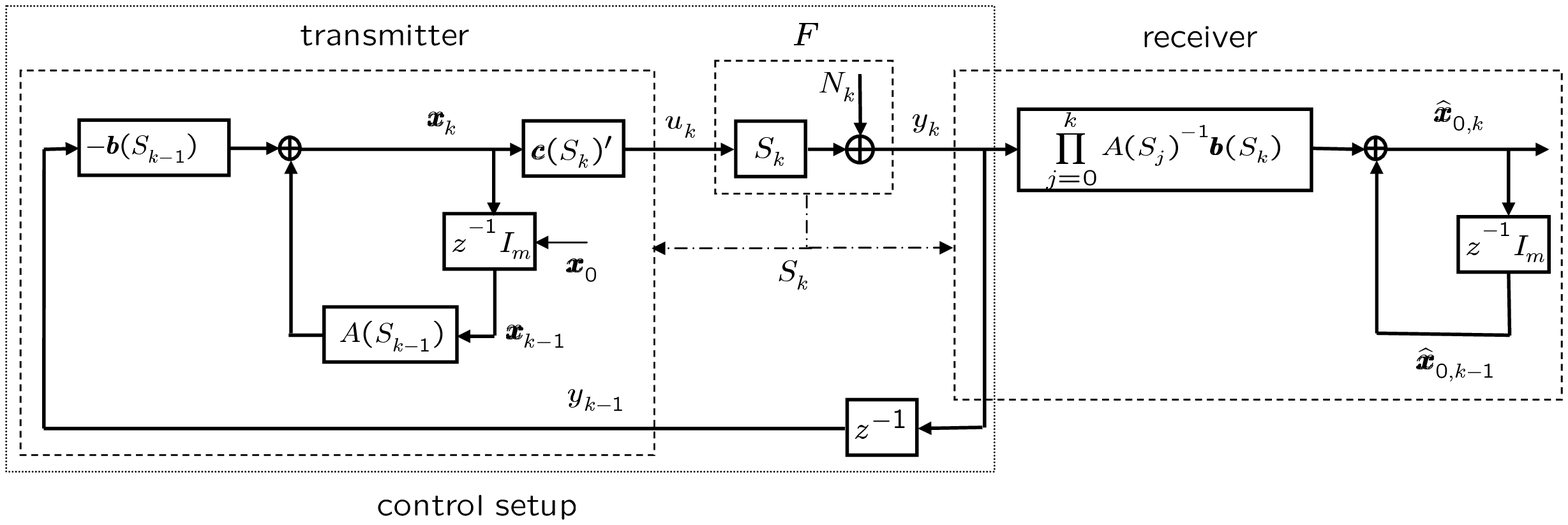}}}
 \caption{The communication system over the channel $\calF_{TCSI}$. The system inside the dotted box represents a closed-loop
control system and will be discussed in Sec. \ref{sec:control}. }
\label{fig:comd0}
\end{center}
\end{figure}

Fig. \ref{fig:comd0} shows the proposed communication system.  Parameters $A \in \mathbb{R}^{m \times m}$, $\bfb \in
\mathbb{R}^{m}$, and $\bfc \in \mathbb{R}^{m}$ depend causally on
the channel states and will be specified shortly.  At time $k$, $k\geq 0$, the system generates signals according to the following
dynamics in the listed order:
\begin{equation} \label{dyn:commud0}
\begin{array}{lll}
\bfx_k &=& A(S_{k-1}) \bfx_{k-1}-\bfb(S_{k-1}) y_{k-1} \\
u_{k} &=& \bfc (S_{k}) ' \bfx_{k}  \\
y_{k} &=& S_{k} u_{k} + N_{k}\\
\hbfx_{0,k} &=& \displaystyle \hbfx_{0,k-1} + \left( \prod_{j=0}^k A(S_{j})^{-1} \right)
\bfb(S_{k}) y_k ,
\end{array}
\end{equation}
where $S_{-1}:=s[1]$, $y_{-1}:=0$, $\bfx_{-1}:=A(S_{-1})^{-1} \bfx_0$, and $\hbfx_{0,-1}:=0$. The
above recursions will generate a sequence of receiver estimates
$\{\hbfx_{0,k}\}$.

%

\subsection{Choice of parameters} \label{d0:choice}

Supposing
that $S_{k-1}=s[j]$ and $S_{k}=s[l]$, we define
\begin{equation}
\begin{array}{llll}
A(S_{k-1})&:= \diag( [1,\cdots,1, a(S_{k-1}),1,\cdots,1] ) & \in \mathbb{R}^{m \times m}\\
\bfb(S_{k-1})&:= [ 0,  \cdots, 0,b(S_{k-1}), 0,\cdots, 0 ]' & \in
\mathbb{R}^{m} \\
\bfc(S_{k})&:= [0,\cdots,0,c(S_{k}),0,\cdots,0]' & \in \mathbb{R}^{m},
 \end{array}\label{eq:choiced0}
 \end{equation}
where $a(S_{k-1})$ is the $(j,j)$th element of $A(S_{k-1})$, given by
\begin{equation}   a(S_{k-1}) : =\sqrt{(S_{k-1})^2 \Gamma(S_{k-1}) +1}
\quad ;\label{d0:choicea}\end{equation}
$b(S_{k-1})$ is the $j$th element of $\bfb(S_{k-1})$, given by
\begin{equation}  b(S_{k-1}) := \frac{\Gamma(S_{k-1}) S_{k-1}
}{a(S_{k-1})}; \label{d0:choiceb} \end{equation}
and $c(S_{k})$ is the $l$th element of
$\bfc(S_{k})$, given by
\begin{equation} c(S_{k}): =1. \label{d0:choicec}\end{equation}

From the above choices one can see that the current CSI ($S_{k}=s[l]$) determines which subsystem ($\Sigma_l$) is selected to use the forward-link channel and hence determines the current transmission power (approximately equal to $\Gamma(s[l])$), and the delayed CSI ($S_{k-1}=s[j]$) determines which subsystem ($\Sigma_j$) is selected to incorporate the delayed output feedback.

\subsection{Encoding and decoding} \label{d0:encdec}

Fix $K$ and $\epsilon>0$.  We define the codebook $v_K$ for the $(M_K,K+1)$ code in the space $\mathbb{R}^m$ such that along each of the $m$ dimensions, the codebook is similar to that for the AWGN channel case. More specifically, let $v_K:= v_K^{(1)} \times v_K^{(2)} \times \cdots v_K^{(m)} $, where $\times$ denotes the Cartesian product, $v_K^{(j)}:= \{ -\sqrt{\Gamma(s_j)} + 2(i-1)D_K^{(j)} , i=1,\cdots,M_K^{(j)}\}$, and
\begin{equation} 
\ba{lll} M_{K}^{(j)}&:=& \disp \lfloor \bara[j] ^{(K+1)(1-\epsilon)}  \rfloor \\
D_K^{(j)}&:=& \disp \frac{\sqrt{\Gamma(s_j)}}{M_{K}^{(j)} -1} .\ea \label{eq:mkjd0}\end{equation}
Note that we define $D_{K}^{(j)}:=0$ if $M_{K}^{(j)}=1$ (which by Lemma \ref{lemma:bara} is equivalent to $\Gamma(s_j)=0$).  Then let
\begin{equation} M_K:=\prod_{j=1}^m  M_K^{(j)} . \end{equation}
That is, $v_K$ contains $M_K$ codewords and each codeword is an $m$-dimensional vector.

\begin{figure}[t!]
\center \scalebox{.49}{\includegraphics{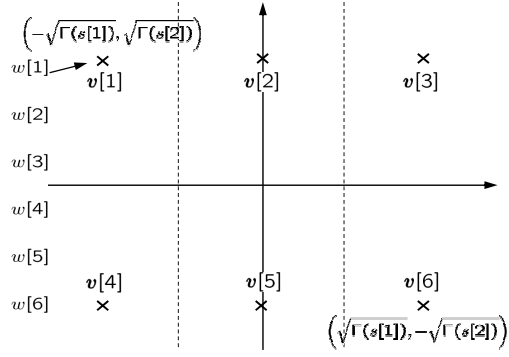}} \caption{ An
example of a codebook.  Assume $m=2$, $M_{K}^{(1)}
=3$, and $M_{K}^{(2)}=2$, namely $M_K=6$. The decision boundaries are the dotted lines and the horizontal axis, which form six decision regions, one for each codeword.  Suppose that the message $w[1]$ is
to be conveyed. Then the codeword $\pmb{v}[1]$ is to be transmitted, and the
two values $-\sqrt{\Gamma(s[1])}$ and $\sqrt{\Gamma(s[2])}$ are to be transmitted through two constant-gain channels, respectively. At
the receiver side, if $\hbfx_{0,K}$ lies in the upper left decision region, then $\pmb{v}[1]$ and hence $w[1]$ can be
correctly recovered.  } \label{fig:hypercube}
\end{figure}

For encoding, suppose $\pmb{V}_K$ is the codeword corresponding to the randomly selected message $W_K$. Let $\bfx_0:=\pmb{V}_K$ which enters the system
(\ref{dyn:commud0}) as the initial condition and will generates the channel
input sequence $u_0^K$. For decoding, based on the channel output sequence
$y_0^K$, the receiver calculates
$\hat{\bfx}_{0,K}$, and then decides $\hat{\pmb{V}}_K$, the codeword closest to $\hbfx_{0,K}$, to be the one
transmitted by the transmitter (closest in the sense of the Euclidean distance).  The decoded message $\hat{W}_K$ can then be obtained. See Fig. \ref{fig:hypercube} for a
simple example of a codebook.

\subsection{System dynamics} \label{d0:decoupling}

We will show that the $m$ subsystems are decoupled from each other, and each of them is are running over a constant-gain channels with output feedback.

\textbf{The closed-loop equation}

We can rewrite the dynamics of $\bfx_k$ in (\ref{dyn:commud0}) as
\begin{equation} \bfx_{k} = A_{cl}(S_{k-1}) \bfx_{k-1} - \bfb(S_{k-1}) N_{k-1},
\label{dyn:xd0} \end{equation}
where
\begin{equation} A_{cl}(S_{k-1}):= A(S_{k-1}) - S_{k-1} \bfb(S_{k-1})
\bfc(S_{k-1})' \label{eq:acld0}\end{equation}
is the closed-loop matrix for generating $\bfx_{k}$.  One can then show that
\begin{equation} A_{cl}(S_{k-1})= A(S_{k-1})^{-1} .\end{equation}
To see this, assume $S_{k-1}=s[j]$.  With our choice of parameters, it holds that
$A_{cl}(S_{k-1}) $ is a \emph{diagonal} matrix whose $(i,i)$th
element is 1 if $i \neq j$, and is
\begin{equation} a(S_{k-1}) - S_{k-1} b(S_{k-1}) c(S_{k-1}) =
a(S_{k-1})^{-1} \label{closeloopad0}\end{equation}
if $i=j$. Hence, we have
\begin{equation} x_{k}^{(i)} =\left \{\ba{lll}   a(S_{k-1})^{-1} x_{k-1}^{(i)} - b(S_{k-1})
N_{k-1}  & \;\; & \myif \;i = j \\
x_{k-1}^{(i)} & \;\; & \myif \;i \neq j   ; \label{dyn:xjd0} \ea  \right . \end{equation}
or equivalently in matrix form
\begin{equation} \ba{lll}\bfx_{k} &=& A(S_{k-1})^{-1} \bfx_{k-1} - \bfb(S_{k-1})
N_{k-1}  . \ea \label{cl_Fd0}\end{equation}

\textbf{The evolution of each subsystem}

Fix any $j$ in $\{1,2,\cdots,m\}$ and the time horizon 0, 1, $\cdots$, $K$.  Now extract the subsequence $\{k_1$,$k_2$,$\cdots$,$k_n\}$ formed by the time instants when the channel state is $s[j]$, viz. $S_{k_i} =s[j]$ for all such $k_i$'s and only such $k_i$'s.
Then at time $k_{i}+1$ the subsystem $\Sigma_j$ updates as
\be \disp x_{k_{i}+1}^{(j)} = a(s[j])^{-1} x_{k_{i}}^{(j)} - b(s[j])
N_{k_{i}} ,\ee
and remains this state value through time $k_{i+1}$, i.e. $x_{k_i +1}^{(j)}=x_{k_{i+1}}^{(j)}$, which will be be updated again at time $k_{i+1}+1$.  Thus, the dynamics of the subsystem $\Sigma_j$ can be equivalently written as one running only on the time instants $k_1,k_2,\cdots,k_n$ and experiencing a flat channel:
\be \disp  x_{k_{i}}^{(j)} = a(s[j])^{-1} x_{k_{i-1}}^{(j)} - b(s[j])
N_{k_{i-1}} . \label{d0:subsysevole} \ee
The value $n$ in the above for the fixed $j$ is 
\be  n:=n(j,k)(S_0^k) := \disp \sum_{t=0}^{k} \pmb{1}(S_t=s[j]) \label{eq:njk} \ee
for $k= 0,1, \cdots, K$, and $\pmb{1}(A)$ is the indicator function which is 1 if $A$ holds true and 0 otherwise. The notation $n(j,k)(S_0^k)$ indicates that $n(j,k)$ is a random variable obtained from the sequence $S_0^k$.  Since $n(j,k)$ is the number of times that the state $s[j]$ is visited during time 0 and time $k$, by ergodicity it holds that
\begin{equation} \disp  \frac{ n(j,k) } {k+1} \arrowP  \pi [j] .
 \label{eq:inprobd0} \end{equation}

\textbf{The end-to-end equation}

\begin{lemma} \label{lemma:end2endd0}
The end-to-end equation is
\be \ba{rl}
\hat{x}_{0,k}^{(j)}
 =& \disp  (1 - a(s[j])^{-2n(j,k)}) x_0^{(j)} +  a(s[j])^{-2n(j,k)}  \sum_{i=1}^{n(j,k)} a(s[j])^{i} b(s[j]) N_{k_i} , \ea \label{eq:hbfxd0}
 \end{equation}
or in matrix form
\begin{equation}
\hat{\bfx}_{0,k}  =\disp  (I - (\Phi_k)^2) \bfx_0 +  ( \Phi_k)^2 \sum_{t=0}^k (\Phi_t)^{-1} \bfb(S_{t}) N_t  ,  \label{eq:hbfxd0m}  \end{equation}
where
\begin{equation}  \disp \Phi_k = \prod_{t=0}^{k}A(S_{t})^{-1} = \disp \diag
\left(\left[ a(s[1])^{-n(1,k)},\cdots, a(s[m])^{-n(m,k)} \right] \right) .
\label{def:Phikd0} \end{equation}
\end{lemma}

\proof Recursively applying the encoder dynamics and closed-loop dynamics one obtains
\begin{equation} \ba{lll}
\bfx_{k+1} &=& \disp \Phi_k \bfx_0 - \Phi_k \sum_{t=0}^k ( \Phi_t )^{-1} \bfb(S_{t}) N_t  \\
&=& \disp (\Phi_k)^{-1} \bfx_0 - (\Phi_k)^{-1} \sum_{t=0}^k  \Phi_t  \bfb(S_{t}) y_t .
\ea  \end{equation}
Then
\begin{equation}
\hat{\bfx}_{0,k}  = \disp \sum_{t=0}^k  \Phi_t  \bfb(S_{t}) y_t= \disp \bfx_0 -\Phi_k \bfx_{k+1}  . \end{equation}
Hence (\ref{eq:hbfxd0m}) follows.  Then by (\ref{def:Phikd0}) and (\ref{d0:subsysevole}), Equation (\ref{eq:hbfxd0}) follows.
 \endproof

\subsection{Coding theorem}

\begin{theorem}\label{th:capd0}
Consider the channel $\calF_{TCSI}$. The communication system
described in (\ref{dyn:commud0}), along with the parameters
given by (\ref{eq:choiced0})-(\ref{d0:choicec}) and encoding/decoding stated in Sec. \ref{d0:encdec}, achieves any rate arbitrarily close to the capacity $C=  \log \tilde{a}$.

\end{theorem}

\proof  The asymptotic signaling rate is
\begin{equation} \ba{lll} R &=& \disp \lim _{K \rightarrow \infty} \frac{\sum_{j=1}^m \log 
M_{K}^{(j)} }{K+1} \\
&=& \disp (1-\epsilon) \sum_{j=1}^m \log \bara[j] \\
& =  & \disp (1-\epsilon) \log \tilde{a}.
\ea \end{equation}

For the average input power, from the decoupling and (\ref{d0:subsysevole}), one can show that the subsystem $\Sigma_i$ has an input power bounded from above by $\Gamma(s[i])$ at any time.  Over all possible channel realizations, $\Sigma_i$ occurs with probability $\pi[i]$.  Since $\sum \pi[i] \Gamma[i] \leq \calP$, the average input power constraint is satisfied. 

We analyze the probability of error in three steps. First, show that it is sufficient to study the behavior of $PE_{K|S}$, i.e. $PE_{K}$ conditioned on the channel state sequence; second, show that it is sufficient to study the behavior of $PE_{K|S}^{(j)}$, namely the conditional probability of error for the $j$th subsystem; and third, analyze $PE_{K|S}^{(j)}$. We define $PE_{K|S}:=\Pr(\hat{\pmb{V}}_K \neq \pmb{V}_K|S_0^K)$ and
\begin{equation} PE_{K|S}^{(j)} : = \Pr \left( \hat{V}_{K}^{(j)} \neq V_K^{(j)} | S_0^{K} \right)
\label{pe:sj}\end{equation}
where $\hat{V}_{K}^{(j)}$ is the $j$th entry of $\hat{\pmb{V}}_K$ and $V_K^{(j)}$ is the $j$th entry of $\pmb{V}_K$. We point out that $PE_{K|S}$ and $PE_{K|S}^{(j)}$ are random variables dependent on $S_0^K$.

Step 1: We will show that $PE_K \rightarrow 0$ holds if $PE_{K|S} \arrowP 0$.  For any $\mu>0$, let
\be \Omega_{K,\mu}:= \left\{ S_0^K  \left|  PE_{K|S} < \mu   \right. \right\}.
\end{equation}
Suppose $PE_{K|S} \arrowP 0$, then there exists $\kappa:=\kappa(\mu)$ such that for any $K>\kappa$, $\Pr (\Omega_{K, \mu}) > 1-\mu$.  Thus, for any $K>\kappa$,
\begin{equation} \ba{lll}
PE_{K} &=& \disp \sum_{S_0^{K} \in \Omega_K} PE_{K|S} \Pr (S_0^{K}) \\
&=& \disp \sum_{S_0^{K} \in \Omega_{K, \mu} } PE_{K|S} \Pr
(S_0^{K}) +
        \sum_{S_0^{K} \not \in \Omega_{K, \mu} } PE_{K|S} \Pr
        (S_0^{K})\\
&< & \disp \sum_{S_0^{K} \in \Omega_{K, \mu} } \mu \Pr
(S_0^{K}) +
        \sum_{S_0^{K} \not \in \Omega_{K, \mu} }  \Pr
        (S_0^{K})\\
&< & \disp \mu  +  (1-\Pr (\Omega_{K, \mu} )) < 2 \mu.
\ea \end{equation}
This implies that $PE_K \rightarrow 0$.

Step 2: Invoking the union bound
\begin{equation} PE_{K|S} = 1- \prod _{j=1}^m (1-PE_{K|S}^{(j)}) \leq
\sum_{j=1}^m PE_{K|S}^{(j)} ,\end{equation}
we conclude that $PE_{K|S} \arrowP 0$ would hold if $PE_{K|S}^{(j)} \arrowP 0$ for all $j$.

Step 3: Now we study $PE_{K|S}^{(j)}$. If $\Gamma(s[j])=0$, i.e. $\bar{a}[j]=1$ (see Lemma \ref{lemma:bara}), then by construction we have $\hatx_{0,K}^{(j)}=x_0^{(j)}=0$ and hence $PE_{K|S}^{(j)}=0$.
Next we focus on the case with $\Gamma(s[j])>0$, i.e. $\bar{a}[j]>1$ and $a(s[j])>1$.
The end-to-end equation (\ref{eq:hbfxd0}) implies that $\hatx_{0,K}^{(j)}$ is Gaussian conditioned on $S_0^K$ and $x_0$:
\begin{equation} \hatx_{0,K|S,\bfx_0}^{(j)} \sim \mathcal{N}\left((1 - a^{-2n}) x_0, \left( a^{-2n} \sqrt{ (1 - a^{-2n}) \calP } \right) ^2 \right)  \end{equation}
where we have defined $n:=n(j,K)$ for convenience. Similar to the case of $\calF_A$, one can derive that
\begin{equation}  PE_{K|S,\bfx_0}^{(j)} \leq \disp 2 Q\left(  \frac{1}{\sqrt{ 1 - a(s[j])^{-2n}}} \left( \frac{a(s[j])^{n}}{ \lfloor \bara[j]^{(K+1)(1-\ep)} \rfloor -1 } -  a(s[j])^{-n}   \right) \right)  . \end{equation}
Since $n \arrowP \infty$ as $K \rightarrow \infty$, it is easily seen that one needs to only show $ \eta_K:= a(s[j])^{-n} \bara[j]^{(K+1)(1-\ep)} \arrowP 0$. However, it holds that
\be \ba{lll} \disp (\eta_K)^{\frac{1}{K+1}}  &=& \disp  a(s[j]) ^ { -\frac{n}{K+1} + \pi[j](1-\ep) }  \\
&\arrowP& \disp  a(s[j])^{ - \pi[j] \ep  } < 1 , \ea \ee
which implies that $PE_{K|S,\bfx_0}^{(j)} \arrowP 0 $, $PE_{K|S}^{(j)} \arrowP 0 $,  and $PE_K \rightarrow 0$.  Note that we have used properties of the convergence in probability; see the Continuous Mapping Theorem and Corollary 3.5 in \cite{book:olav}.  \endproof

\section{The optimal scheme for the channel $\calF_{I,DTCSI}$}
\label{sec:iid}

In this section we will present an \emph{adaptive scalar system without power adaptation or multiplexing} and show it is optimal for the channel $\calF_{I,DTCSI}$.

For the channel $\calF_{I,DTCSI}$, the DTCSI cannot be used by the transmitter to infer and adapt to future channel states, since the DTCSI is independent of the future channel states.  Nevertheless, the DTCSI can be used by the transmitter to process the delayed output feedback in a way matching the receiver's last operation which was adapted to the instantaneous CSI.

Similar to the case without output feedback, a fixed transmission power is to be used at all times.  Without the need for power adaptation, one can design an optimal system without resorting to multiplexing, an observation made in \cite{shamai_csi99}.  Thus, one can design a one-dimensional but time-varying scheme to adapt to \emph{any channel state} (and hence not necessarily a finite number of state values).  In what follows we briefly introduce the infinite-state channels and present the optimal scheme.

\subsection{The Gaussian i.i.d. fading channel with possibly infinite channel states}

Gaussian i.i.d. fading channels with possibly infinite channel states include
many channels as special cases, such as the Rayleigh, Rician,
Nakagami, and Weibull fading channels. Assume that the
channel states form a discrete-time i.i.d. process with density $p_S(s)$ and that the first and second moments exist.  Denote the corresponding interconnected channel with DTCSI as $\calF_{II,DTCSI}$. Following the steps used to establish the capacities in Section \ref{subsec:cap}, one can show that the channel capacity is given by
\begin{equation} C(\calP)=\frac{1}{2} \E_{S \sim p_S}\log (1+S^2
\mathcal{P}) = \log \tilde{a}  \label{eq:ciidinf}\end{equation}
where $\tilde{a}:=\exp( \E \log a(S) )$ and $a(S):=\sqrt{1+S^2 \mathcal{P}}$.

\subsection{The proposed communication system} \label{subsec:iid}

At time $k$, $k\geq 0$, the system generates signals according to the following
dynamics in the listed order:
\begin{equation} \label{dyn:commuiid}
\begin{array}{lll}
x_k &=& a(S_{k-1}) x_{k-1}-b(S_{k-1}) y_{k-1} \\
u_{k} &=& x_{k}  \\
y_{k} &=& S_{k} u_{k} + N_{k}\\
\hat{x}_{0,k} &=& \displaystyle \hat{x}_{0,k-1} + \left( \prod_{t=0}^k a(S_{t})^{-1} \right) b(S_{k}) y_k ,
\end{array}
\end{equation}
where $S_{-1}:=0$ (or any given number that $S$ can be), $y_{-1}:=0$, $x_{-1}:=a(S_{-1})^{-1} x_0$, and $x_{0,-1}:=0$.  The parameters are
\begin{equation}
\begin{array}{rl}
a(S_{k-1}) & \disp:=\sqrt{(S_{k-1})^2 \mathcal{P} +1}  \\
b(S_{k-1}) & \disp:= \frac{S_{k-1}\mathcal{P}}{a(S_{k-1})} .
 \end{array}\label{eq:choiceiid}
 \end{equation}
The encoding and decoding processes are the same as those for the channel $\calF_C$, except that $M_K$ is now defined as
\begin{equation} M_K := \exp \left( (K+1)(1-\ep) \log \tilde{a} \right)  . \label{MK:iid}\end{equation}
The closed-loop system evolves according to
\begin{equation} x_k = a(S_{k-1})^{-1} x_{k-1} - b(S_{k-1}) N_{k-1} . \end{equation}
Let
\be \phi_k:=\prod_{t=0}^{k} a(S_{t})^{-1} ,\ee
then it holds that
\begin{equation} \ba{lll}
x_{k+1} &=& \disp \phi_k x_0 - \phi_k \sum_{t=0}^k ( \phi_t )^{-1} b(S_{t}) N_t  \\
&=& \disp (\phi_k)^{-1} x_0 - (\phi_k)^{-1} \sum_{t=0}^k  \phi_t  b(S_{t}) y_t .
\ea \label{eq:xiid}  \end{equation}
Hence the end-to-end equation is
\be
\hat{x}_{0,k}
 =\disp  (1 - (\phi_k)^2) x_0 +  ( \phi_k)^2 \sum_{t=0}^k (\phi_t)^{-1} b(S_{t}) N_t  . \label{eq:hbfxiid}
 \end{equation}

\subsection{Coding theorem}

\begin{theorem}\label{th:capiid}
Consider the channel $\calF_{II,DTCSI}$. The communication system
described in (\ref{dyn:commuiid}), along with the parameters
given by (\ref{eq:choiceiid}) and encoding/decoding stated in Sec. \ref{subsec:iid}, achieves any rate arbitrarily close to the capacity $C = \log \tilde{a} $.
\end{theorem}

\proof  See Appendix \ref{app:inf}.
\endproof

\section{The optimal scheme for the channel $\calF_{DTCSI}$} \label{sec:fsmc}

In this section, we present the optimal feedback communication scheme for the channel $\calF_{DTCSI}$ in which the forward link is a generic FSMC.  It is a \emph{multiplexed adaptive system with power adaptation and with augmented channel states}. The main idea behind the scheme is described as follows. Suppose that $m$ subsystems are constructed for $\calF_{DTCSI}$ and to be multiplexed.  Under what condition should the subsystem $\Sigma_j$ transmit over the forward link at time $k$?  Of course any information about $S_k$ cannot be used.  A logic choice is that $\Sigma_j$ transmits over the forward link if and only if $S_{k-1}=s[j]$, viz. the transmitter utilizes the most recent CSI available.  This also leads to the power adaptation based on the DTCSI, which is needed to achieve the capacity as mentioned before.  However, this means that $\Sigma_j$ does not experience a constant-gain channel as it does in the TCSI case.  Consequently, the receiver needs to adapt to both $S_{k-1}$ (to match the transmitter's operation) and $S_k$ (to account for the channel state at time $k$).  In other words, an augmented channel state $(S_{k-1},S_{k})$ is needed at the receiver at time $k$ and therefore, an augmented channel state $(S_{k-2},S_{k-1})$ (which is a delayed version of the one used at the receiver) is needed at the transmitter at time $k$.

\subsection{The proposed communication system} \label{sub:commscheme}

\begin{figure}[h!]
\begin{center}
{\scalebox{.65}{\includegraphics{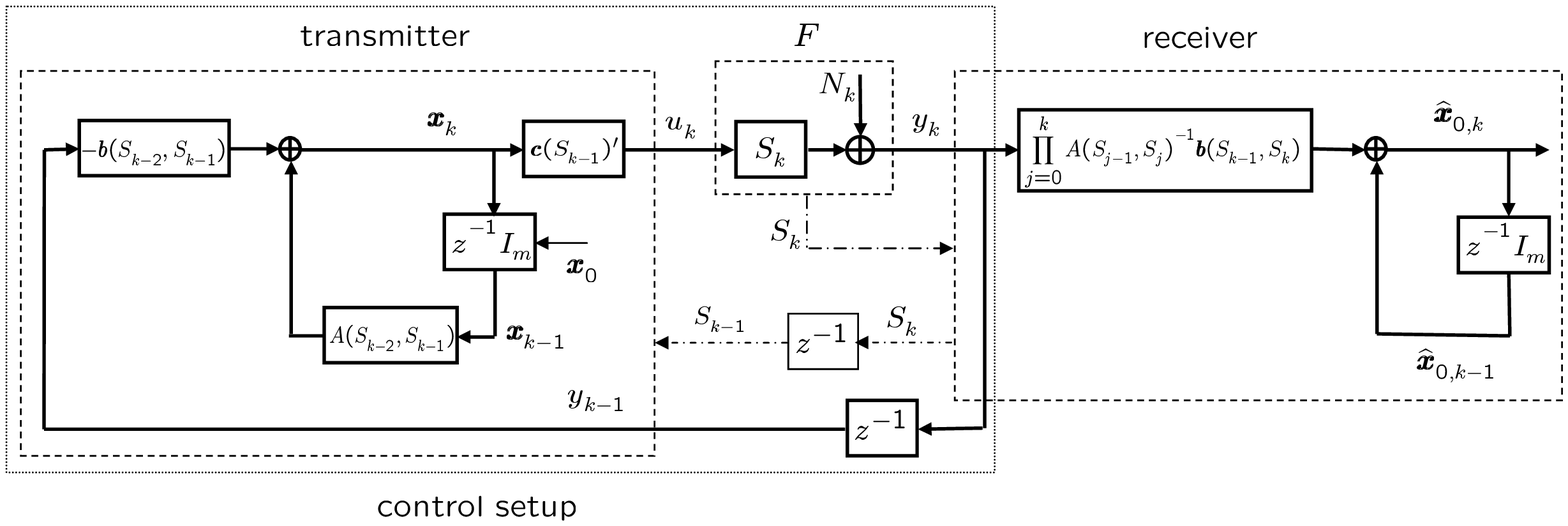}}}
 \caption{The communication scheme for the channel $\calF_{DTCSI}$. The system inside the dotted box represents a closed-loop
control system and will be discussed in Sec. \ref{sec:control}.}
\label{fig:commu}
\end{center}
\end{figure}

Fig. \ref{fig:commu} shows the proposed communication scheme.
At time $k$, $k\geq 0$, the system generates signals according to the following
dynamics in the listed order:
\begin{equation} \label{dyn:commu}
\begin{array}{rll}
\bfx_k &=& A(S_{k-2},S_{k-1}) \bfx_{k-1}-\bfb(S_{k-2},S_{k-1}) y_{k-1} \\
u_{k} &=& \bfc (S_{k-1}) ' \bfx_{k}  \\
y_{k} &=& S_{k} u_{k} + N_{k}\\
\hbfx_{0,k} &=& \displaystyle \hbfx_{0,k-1} + \left( \prod_{j=0}^k A(S_{j-1},S_{j})^{-1} \right)
\bfb(S_{k-1},S_{k}) y_k ,
\end{array}
\end{equation}
where $S_{-2}:=s[1]$, $S_{-1}:=s[1]$, $y_{-1}:=0$, $\bfx_{-1}:=A(S_{-2},S_{-1})^{-1} \bfx_0$, and $\hbfx_{0,-1}:=0$.
%

\subsection{Choice of parameters} \label{subsec:choice}

Supposing
that $S_{k-2}=s[j]$ and $S_{k-1}=s[l]$, we define
\begin{equation}
\begin{array}{llll}
A(S_{k-2},S_{k-1})&:= \diag( [1,\cdots,1, a(S_{k-2},S_{k-1}),1,\cdots,1] ) & \in \mathbb{R}^{m \times m}\\
\bfb(S_{k-2},S_{k-1})&:= [ 0,  \cdots, 0,b(S_{k-2},S_{k-1}), 0,\cdots, 0 ]' & \in
\mathbb{R}^{m} \\
\bfc(S_{k-1})&:= [0,\cdots,0,c(S_{k-1}),0,\cdots,0]' & \in \mathbb{R}^{m},
 \end{array}\label{eq:choice}
 \end{equation}
where $a(S_{k-2},S_{k-1})$ is the $(j,j)$th element of $A(S_{k-2},S_{k-1})$, given by
\begin{equation}   a(S_{k-2},S_{k-1}) : =\sqrt{(S_{k-1})^2 \Gamma(S_{k-2}) +1}
\quad ;\label{eq:choicea}\end{equation}
$b(S_{k-2},S_{k-1})$ is the $j$th element of $\bfb(S_{k-2},S_{k-1})$, given by
\begin{equation}  b(S_{k-2},S_{k-1}) := \frac{ S_{k-1}\Gamma(S_{k-2})}{a(S_{k-2},S_{k-1})} ; \label{eq:choiceb} \end{equation}
and $c(S_{k-1})$ is the $l$th element of
$\bfc(S_{k-1})$, given by
\begin{equation} c(S_{k-1}): =1. \label{eq:choicec}\end{equation}
Whenever $S_{k}$, $k<0$, is encountered, it is treated as $s[1]$.  Note that the above choice of $A$ and $\bfb$ uses the
augmented channel state $(S_{k-2},S_{k-1})$.

The encoding/decoding method and parameters are the same as those presented for the channel $\calF_{TCSI}$ in Sec. \ref{d0:encdec}.  (Of course when computing $M_K^{(j)}$ the expression of $\bar{a}[j]$ for $\calF_{DTCSI}$ as given in (\ref{eq:abar_d1}) should be used instead of (\ref{eq:abar_d0}).)


\subsection{The closed-loop equation and end-to-end equation} \label{subsec:decoup}

The closed-loop dynamics is
\begin{equation} \bfx_{k} = A_{cl}(S_{k-2},S_{k-1}) \bfx_{k-1} - \bfb(S_{k-2},S_{k-1}) N_{k-1},
\label{dyn:x} \end{equation}
where
\begin{equation} A_{cl}(S_{k-2},S_{k-1}):= A(S_{k-2},S_{k-1}) - S_{k-1} \bfb(S_{k-2},S_{k-1})
\bfc(S_{k-2})' .\label{eq:acl}\end{equation}
One can again show that $A_{cl}(S_{k-2},S_{k-1}) =
A(S_{k-2},S_{k-1})^{-1}$, and hence
\begin{equation} \ba{lll}\bfx_{k} &=& A(S_{k-2},S_{k-1})^{-1} \bfx_{k-1} - \bfb(S_{k-2},S_{k-1})
N_{k-1}  . \ea \label{dyn:closechoice}\end{equation}
%

Similar to the case for $\calF_{DTCSI}$, the end-to-end equation can be shown to be
\begin{equation}
\ba{lll} 
\hat{\bfx}_{0,k} &=& \disp \bfx_0 -\Phi_k \bfx_{k+1} \\
 &=&\disp  (I - (\Phi_k)^2) \bfx_0 +  ( \Phi_k)^2 \sum_{t=0}^k (\Phi_t)^{-1} \bfb(S_{t-1},S_{t}) N_t     \\
\hat{x}_{0,k}^{(j)}
 &=&\disp  (1 - (\phi_k^{(j)})^2) x_0^{(j)} +  ( \phi_k^{(j)})^2  \sum_{S_{t-1}=s[j],t \in \{0,...,k\}} (\phi_t^{(j)})^{-1} b(s[j],S_{t}) N_t  , \label{eq:hbfx}
  \ea \end{equation}
where
\begin{equation} \ba{lll} \disp \Phi_k &:=& \disp \diag\left(\left[\phi_k^{(1)},\cdots,\phi_k^{(m)}\right] \right) := \prod_{t=0}^{k}A(S_{t-1},S_t)^{-1} \\
&=& \disp \diag
\left(\left[\prod_{l=1}^{m} a(s[1],s[l]
)^{-n(1,l,k)},\cdots,\prod_{l=1}^{m} a(s[m],s[l] )^{-n(m,l,k)} \right] \right) ; \\
\ea  \label{def:Phik} \end{equation}
in which
\be  n(j,l,k):=n(j,l,k)(S_0^k) := \disp \sum_{t=0}^{k} \pmb{1}(S_{t-1}=s[j],S_{t}=s[l])   \ee
for $j, l = 1, 2, \cdots, m$, and $\pmb{1}(A,B)$ is the indicator function which is 1 if $A$ and $B$ hold true and 0 otherwise.
By ergodicity it holds that
\begin{equation} \disp  \frac{ n(j,l,k) } {k+1}  \arrowP   \pi [j] p_{jl} . \label{eq:inprob_F} \end{equation}
%


The end-to-end equation indicates that each value of
$x_0^{(j)}$ is transmitted independently from other sub-codewords
.

\subsection{Coding Theorem}
\label{sub:fsmccap}

\begin{theorem}\label{th:cap}
Consider the channel $\calF_{DTCSI}$. The communication system
described in (\ref{dyn:commu}), along with the parameters
given by (\ref{eq:choice})-(\ref{eq:choicec}) and encoding/decoding stated in Sec. \ref{d0:encdec}, achieves any rate arbitrarily close to the capacity $C = \log \tilde{a} $.
\end{theorem}

\proof See Appendix \ref{sub:rate}.  \endproof

\section{Connections with feedback control} \label{sec:control}

In the communication schemes discussed above, the closed-loop dynamics can be viewed as \emph{feedback control systems}, which we refer to as the \emph{control setups} associated with the communication systems (the control setups are specified in Fig. \ref{fig:awgn}, Fig. \ref{fig:comd0}, and Fig. \ref{fig:commu} within the dotted boxes).  This naturally draws connections between feedback communication and feedback control.
We note that in the literature there is an increasing trend to explore the intrinsic connections between information theory and control theory, especially when channel output feedback is used in the communication systems \cite{mitter:talk,mitter:markov01,tati:capI,elia_c5,anant_eras04,sahai:main}.  In this section, we will extend some connections between information and control known mainly for linear time-invariant (LTI) systems to systems over FSMCs. In particular, we will see that the optimality in the proposed communication systems can be completely characterized by studying the control setups.

For completeness, we present a
rather brief review of the interactions between information and control. In \cite{tati:capI}, the authors formulated the
feedback capacity problem as a stochastic optimal control
problem, and provided a dynamical programming based solution.
In \cite{sahai:main}, it was revealed the fundamental connections between the
communication of non-stationary, non-ergodic sources and the
stabilization of unstable systems. In \cite{elia_c5}, it was established, over a Gaussian time-invariant channel, the equivalence of feedback
communication and feedback stabilization problems, and that the optimality in
the two problems coincides.  The present paper generalizes mainly along the line of
\cite{elia_c5} to address Gaussian time-varying fading channels.

\subsection{The control setup} \label{sub:control}

We focus on the channel $\calF_{DTCSI}$ unless otherwise specified; other channels can be treated in a similar way or as specializations.  Consider a Markov Jump Linear System (MJLS)
\begin{equation} \label{dyn:commu_mec}
\begin{array}{lll}
\bfx_{k+1} &=& A(S_{k-1},S_{k}) \bfx_{k}-\bfb y_{k} \\
u_{k} &=& \bfc (S_{k-1}) ' \bfx_{k}  \\
y_{k} &=& S_{k} u_{k} + N_{k},
\end{array}
\end{equation}
in which $A(S_{k-1},S_k)$ and $\bfc(S_{k-1})$
are given as in (\ref{eq:choice}), (\ref{eq:choicea}), and (\ref{eq:choicec}).  As before the system state is $\bfx$, which updates according to the first equation in (\ref{dyn:commu_mec}) and is driven by the initial condition $\bfx_0$ and the controller's output $\bfb y_{k}$.  The system's output $u_k$ is linear in the system state $\bfx_k$.  However the controller does not have access to either $u_k$ or $\bfx_k$; it can merely utilize $y_k$, a scaled and noisy measurement of $u_k$, and $\bfb$ is the controller gain.  The goal is to design the controller gain $\bfb$ to ensure closed-loop stability (to be defined), with the discrete state
$S_0^k$ known to the controller when $\bfx_{k+1}$ is computed.  Namely, we wish to stabilize the MJLS based on the corrupted observation $y_{k}$ and perfect knowledge of the Markov state $S_0^{k}$.  Though we are not aware of any reference with an MJLS with our particular choices of $A$ and $\bfc$, this does not prevent us from studying the control of this ``conceptual'' system.

The open loop of this system, namely $\bfx_k = A(S_{k-2},S_{k-1}) \bfx_{k-1}$ (obtained by letting $\bfb:=0$), is unstable and $\bfx_k$ will grow unboundedly since $A(S_{k-1},S_k) \geq 1$.  We can define and compute the average rate of growth of $\bfx_k$ in the open loop as
\begin{equation} \ba{lll}
& & \disp \lim_{k \rightarrow \infty}  \frac{1}{k+1}\log \frac{|\prod_{j=1}^m  x_k^{(j)}|}{|\prod_{j=1}^m  x_0^{(j)}|} \\
&=& \disp \lim_{k \rightarrow \infty}  \frac{1}{k+1}\log \prod_{j=1}^m  (\phi_k^{(j)})^{-1} = \log \tilde{a},
\ea \label{fsmc:DI}
\end{equation}
where the last equality is to be interpreted as convergence in probability. The larger the open-loop growth rate is, the more unstable the open-loop MJLS is considered to be.  Hence the open-loop growth rate can be seen as an indicator of how unstable the open loop is and is the counterpart of the ``degree of instability'' (in log scale) defined for an LTI system in \cite{elia_c5}.

We say the system is \emph{mean-square stabilized} (MSS) if in the closed loop, it holds that $\E \bfx_k \arrowP 0$ and $\E (\bfx_k)^2$ goes to some constant as $k \rightarrow \infty$.   The necessary and sufficient condition for an MJLS to be MSS can be found in \cite{costa:book} (Theorems 3.9 and 3.33).

\subsection{Feedback stabilization implies reliable communication}

If the MJLS, unstable in the open loop, is MSS in the closed loop, then its associated communication system can achieve any rate $R$  arbitrarily close to the open-loop average rate of growth $\log \tilde{a}$.  To see this, suppose $\bfb$ is chosen (not necessarily according to the capacity-achieving choice (\ref{eq:choiceb})) such that the closed-loop dynamics
\be \bfx_{k+1} = A_{cl}(S_0^k) \bfx_{k} - \bfb(S_0^k) N_{k}
 \ee
is MSS, where $A_{cl}(S_0^k):= A(S_{k-1},S_{k}) - S_{k} \bfb(S_0^k)
\bfc(S_{k-1})' $.
%
Define $\hat{\bfx}_{0,k}$ according to (\ref{dyn:commu}), one can again obtain that
\begin{equation}
\hat{\bfx}_{0,k} = \bfx_0 -\Phi_k \bfx_{k+1} , \label{eq:neare2e}\ee
namely, the relation among $\hat{\bfx}_{0,k}$, $\bfx_0$, and $\bfx_{k+1}$ remains invariant for any $\bfb$ (see also (\ref{eq:hbfx}) for the same equation with the capacity-achieving $\bfb$). Since  $\Phi_k$ decays exponentially at rate $\log \tilde{a}$, and since the first and second moments of $\bfx_k$ converge to certain constants due to MSS, this relation implies that the difference between $\hat{\bfx}_{0,k}$ and $\bfx_0$ vanishes exponentially, from which the reliable communication can be concluded if the encoding/decoding process described in Section \ref{subsec:choice} is used.  Indeed, one can derive that
\be \hbfx_{0,k|S,\bfx_0}^{(j)} \sim \mathcal{N} \left( (1 - \phi_k^{(j)} \psi_k^{(j)} ) x_0^{(j)}, (\phi_k^{(j)} \sigma_k^{(j)} )^2    \right) , \ee
%
where $\psi_k^{(j)}:=\E (x_k^{(j)} | S_0^k)$ and $\sigma_k^{(j)}:= \sqrt{\E (x_k^{(j)}| S_0^k)^2} $.  By the MSS, it holds that $\psi_k^{(j)} \arrowP 0$ and $\sigma_k^{(j)}$ converges to some constant in probability.
%
%
One can show the probability of error satisfies
\begin{equation}  PE_{K|S,\bfx_0}^{(j)} \leq  \disp 2 Q\left( \frac{ D_K^{(j)} - \phi_k^{(j)} \psi_k^{(j)} |x_0^{(j)}| }{\phi_k^{(j)} \sigma_k^{(j)}}\right)  . \ee
It then suffices to show $\phi_k^{(j)}/D_K^{(j)} \arrowP 0$, which is indeed true as we have proven before.  Thus the closed-loop stability implies that the corresponding communication system can transmit reliably at rate $R=(1-\ep) \log \tilde{a}$.

Several remarks follow.  First, our proofs of vanishing probability of error presented before are in essence based on analyzing the closed-loop system dynamics and their asymptotic behavior that are in fact stabilization analysis. It is therefore not surprising to see a general connection between stabilization and reliable communication exists for those systems. Second, the communication rate $R$ is determined by the open-loop growth rate (and the small slack $\ep$) but \emph{independent} of how $\bfb(S_0^K)$ is chosen as long as it stabilizes the closed loop, a property to be further explored in the next subsection.  Third, one can easily verify that our choice of $\bfb$ in (\ref{eq:choiceb}) indeed leads to MSS of the closed loop, which immediately leads to the conclusion of reliable communication of the proposed system.

\subsection{Cheap control}

Since an arbitrary stabilizing $\bfb(S_0^K)$ guarantees reliable communication at rate $(1-\ep) \log \tilde{a}$, we can select a stabilizing $\bfb(S_0^K)$ to minimize
the power of $u$, namely the transmission power.  This is a control problem known as the cheap control problem; see \cite{mincontrol:book,elia_c5} for the LTI formulation of cheap control.  Precisely,  we need to solve the following optimal control problem over the MJLS:
\begin{equation} \ba{lcl}   &\displaystyle \min  _{\bfb  \textnormal{ \small}} & \disp
  \frac{1}{K+1}\E \sum_{k=0}^K (u_k)^2 , \ea \label{opt:fsmc:mec}\end{equation}
in which there is no direct penalty on the control effort $b y_k$; hence the name ``cheap control'' \footnote{The reader may find in some references (e.g. \cite{elia_c5}) the minimization of the transmission power in a communication system is transformed into a control problem called the \emph{expensive control} as opposed to cheap control. We remark that the cheap control problem and expensive control problem can be reformulated as one another, depending on whether one treats $b$ or $c$ as the controller gain, and in either case the optimal controller places the closed-loop eigenvalues at the reciprocals of open-loop eigenvalues.  Specifically in expensive control, one views $u$ as the controller's output, $c$ as the state-feedback controller gain to be designed, and $b$ as given, and one needs to minimize the power of the control effort subject to closed-loop stability.}.  The minimum power obtained from the solution to the cheap control problem is equal to the optimal transmission power in the corresponding communication system, and the optimizing $\bfb$ is the one given in (\ref{eq:choiceb}), which is readily shown using proof by contradiction.
It is well known that the solution to cheap control over an LTI system is such that the closed-loop eigenvalues are placed at the reciprocal locations of the open-loop eigenvalues.  This is still the case for the cheap control over the MJLS studied in this paper since all of our proposed communication schemes are such that $A_{cl}=A^{-1}$.

\subsection{The control-oriented approach}

Combining the control-oriented characterizations of both the achievable communication rate and transmission power, we conclude that the optimality in the communication systems coincides with that in the control setups, and if one solves the cheap control for the appropriately formulated MJLS, then the capacity-achieving coding scheme can be obtained. Since investigating the control problem does not require notions such as the transmitter, receiver, codebooks, encoding/decoding, and probability of error that are present in the communication problem, and since the probability of error analysis is essentially a stability analysis, in certain cases one may choose to first study the control setup and then transform the obtained optimal control system to the optimal communication system.
This was the approach adopted in developing our schemes (despite the fact that the schemes may also be conceived, derived, and presented in a purely information theoretic fashion) and we will briefly discuss this approach below.

The following steps were adopted in constructing the schemes achieving the capacity $C(\calP)$ for the channel $y_k=S_k u_k +N_k$; whether and how they may be extended to more general feedback communication problems (e.g. MIMO problems) remain to be seen. 1) Construct an open-loop unstable MJLS such that the open-loop growth rate is equal to $C(\calP)$.   All open-loop eigenvalues should be outside or on the unit circle. 2) Close the loop over the channel and place the closed-loop eigenvalues at the reciprocal locations of the open-loop eigenvalues; thus the closed loop does not have any unstable eigenvalues.  Then the closed-loop MSS will follow if the eigenvalues on the unit circle do not occur with probability one. 3) Verify the average power of $u$ is no greater than $\calP$.  4) Add an equation to recover the initial condition of the control system which effectively transforms the control system to a feedback communication system.  Some examples follow.

\textbf{Example: $\calF_{I,DTCSI}$}

1) The capacity expression (\ref{C_d0d1}) suggests a system with open-loop eigenvalues $a(s[j])$, $j=1,\cdots,m$, where $a(s[j])$ is defined in (\ref{eq:abar_iid}) and satisfies $|a(s[j])|\geq 1$.  If the eigenvalue $a(s[j])$ occurs whenever $S_{k-1}=s[j]$, then the open-loop growth rate equals the capacity rate (since $C(\calP)=\sum \pi[j] \log a(s[j])$), and the unit-circle eigenvalues do not occur with probability one.  Therefore, the open-loop MJLS may be either a scalar system $x_k=a(S_{k-1}) x_{k-1}$, or a multiplexed system with the $j$th subsystem being $x_k^{(j)}=a(S_{k-1}) x_{k-1}^{(j)}$ activated when $S_{k-1}=s[j]$.

2) If the scalar open-loop system is considered, then the system with control input is $x_k=a(S_{k-1}) x_{k-1} + b y_k$ and the controller $b$ is to be specified.  One can choose $b:=b(S_{k-1})$ according to (\ref{eq:choiceiid}), which leads to the closed-loop dynamics $x_k = a(S_{k-1})^{-1} x_{k-1} - b(S_{k-1}) N_{k-1}$ (i.e. the closed-loop eigenvalue is the reciprocal of the open-loop one) and is MSS.
Likewise, one can see the same choice of $b(S_{k-1})$ places the closed-loop eigenvalues of the multiplexed system at the reciprocal locations of open-loop ones and hence leads to MSS. Thus for either construction the closed loop is MSS and any rate arbitrarily close to $C(\calP)$ is achievable.

3) The average power can be verified directly for either construction.

4) Recover $x_0$ from $y_0^K$.  In the scalar system case, this can be done by using a smoothed estimator, or simply, by setting $\hat{x}_{0,k}  :=  \sum_{t=0}^k  \phi_t  b(S_{t}) y_t  $
since this leads to $\hat{\bfx}_{0,k}  = x_0 - \phi_k x_{k+1}$.  That is, the difference between $\hat{x}_{0,k}$ and $x_0$ vanishes exponentially.  The multiplexed system case can be dealt with similarly.

Therefore, one can construct either a scalar system or a multiplexed system to achieve $C(\calP)$ for $\calF_{I,DTCSI}$.  In addition, using this approach, one can also see that the scalar system cannot achieve the capacity for $\calF_{DTCSI}$ in general (in the third step, verifying the power would fail), but the multiplexed system can.  It is also evident from this approach that $\calF_{DTCSI}$ in general requires the augmented channel states to be used in the optimal scheme: since the capacity expression (\ref{C_d0d1}) uses two channel states $S_{-d}$ and $S$ for $\calF_{DTCSI}$, each open-loop eigenvalue needs to depend on two channel states.

It is intriguing to ask under what condition a scalar system can achieve the capacity of a channel. For the general FSMC defined in this paper, we have seen that if a uniform power allocation is suggested by the capacity expression, a scalar system without multiplexing can achieve the capacity; otherwise a multi-dimensional system with multiplexing needs to be used to achieve the capacity.

\textbf{Example: an FSMC with multi-step delayed feedback}

Consider an FSMC with both the CSI and channel output feedback delayed by $d \geq 1$ steps at the transmitter.   1) Design the open-loop MJLS such that the subsystem $\Sigma_j$ is activated at time $k$ if and only if $S_{k-2d}=s[j]$, and when activated the subsystem evolves as
\be x_k^{(j)}=a(S_{k-2d},S_{k-d}) x_{k-d}^{(j)}, \label{dyn:anyd} \ee
where $a(S_{k-2d},S_{k-d}):=\sqrt{(S_{k-d})^2 \Gamma(S_{k-2d})+1}$.  This results in that the open-loop growth rate is equal to $C(\calP)=\E \log a(S_{k-2d},S_{k-d}) $. Equation (\ref{dyn:anyd}) implies that $d$ initial condition values need to be specified to completely define the subsystem dynamics, namely  $x_0^{(j)},\cdots,x_d^{(j)}$ need to be charged instead of being generated on the fly from the dynamics. Consequently, the initial condition of the MJLS needs to specify totally $dm$ values, which translates into a codebook with each codeword being a $dm$-dimensional vector. 2) Let $b(S_{k-2d},S_{k-d}) :=  S_{k-d}\Gamma(S_{k-2d}) / a(S_{k-2d},S_{k-d})$ and then the closed-loop eigenvalue is $a(S_{k-2d},S_{k-d})^{-1}$, and thus MSS follows.  Then the steps 3) to 4) are rather straightforward and the detail is skipped.

To summarize, we have seen that the control-oriented approach is a powerful tool in studying the feedback communication problems.

\section{A numerical example} \label{sec:eg}

Consider a Gilbert-Elliot fading channel with DTCSI, output feedback, and AWGN, i.e. an $\calF_{DTCSI}$ with $m=2$; see Fig. \ref{fig:simu} (a) for the channel state transitions. We simulate the proposed scheme for this
channel.  Fig. \ref{fig:simu} (b) shows the simulated
$PE_{K|S}^{(j)}$ and $PE_{K|S}$ for a randomly chosen sequence $S_0^{19}$, as
well as $PE_{K|S}$ computed using the exact analytic expression. We see that simulated $PE_{K|S}$ decays rather
fast within 20 channel uses and is consistent with the theoretic $PE_{K|S}$. However, the decay of $PE_{K|S}^{(j)}$ and $PE_{K|S}$ is not quite smooth, caused by instantaneous deviations from
the typical channel state behavior (namely, $(n(j,l,K) / (K+1) -  \pi[j] p_{jl})$ may fluctuate considerably around zero).  This may be improved by considering a ``turbo mode" of
using larger power at the moments of large instantaneous
deviations from the typical state behavior, which does not affect the
average power constraint
\cite{sahai_ge05}. Fig. \ref{fig:simu}
(c) shows the decay of ${PE}_k$, where $\epsilon>0$ is the slack from the capacity $C$. In Fig. \ref{fig:simu} (d) we compare the transmitted message and the decoded message bit by bit and
count how many bits are correctly obtained by the receiver. For
$K=24$, the channel can transmit $35.8$ bits if at each step the
capacity $C$ is attained, and the simulation shows that on average
$34.9$ bits are actually correctly decoded.  It would be interesting to compare the bit error rate performance and frame error rate performance (which are related but not identical to the probability of error) of our feedback scheme with the schemes based on capacity-approaching codes such as LDPC codes but without output feedback; however,
to define and perform a fair and accurate comparison is beyond the main scope of this paper and is subject to future work.

\begin{figure}[ht!]
\center \subfigure[
]{\scalebox{.45}{\includegraphics{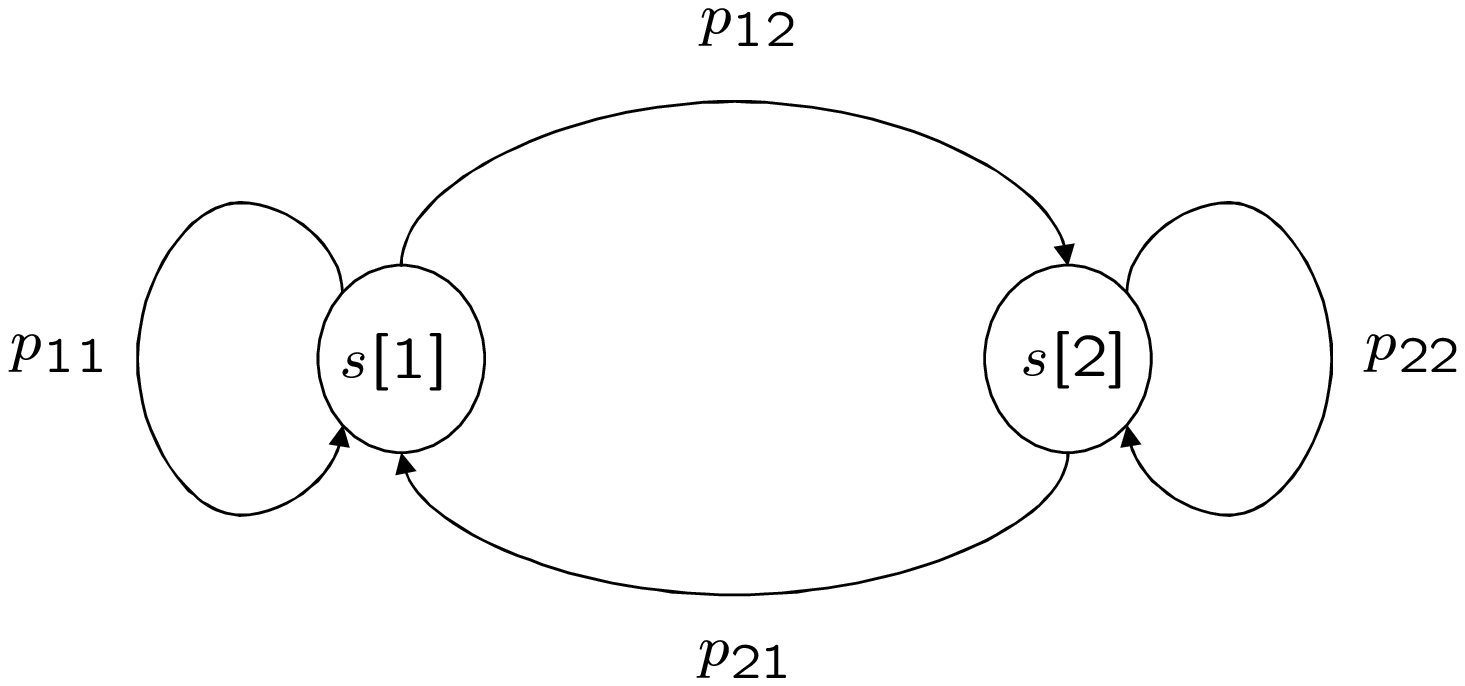}} }
\psfragscanon \psfrag{k}{\LARGE \bf $(K+1)$}  \psfrag{alpha}{ $PE_{K|S}^{(1)}$}
\psfrag{beta}{ $PE_{K|S}^{(2)}$}  \psfrag{gamma}{ $PE_{K|S}$}
\hspace{-0pt}
\subfigure[
]{\scalebox{.52}{\includegraphics{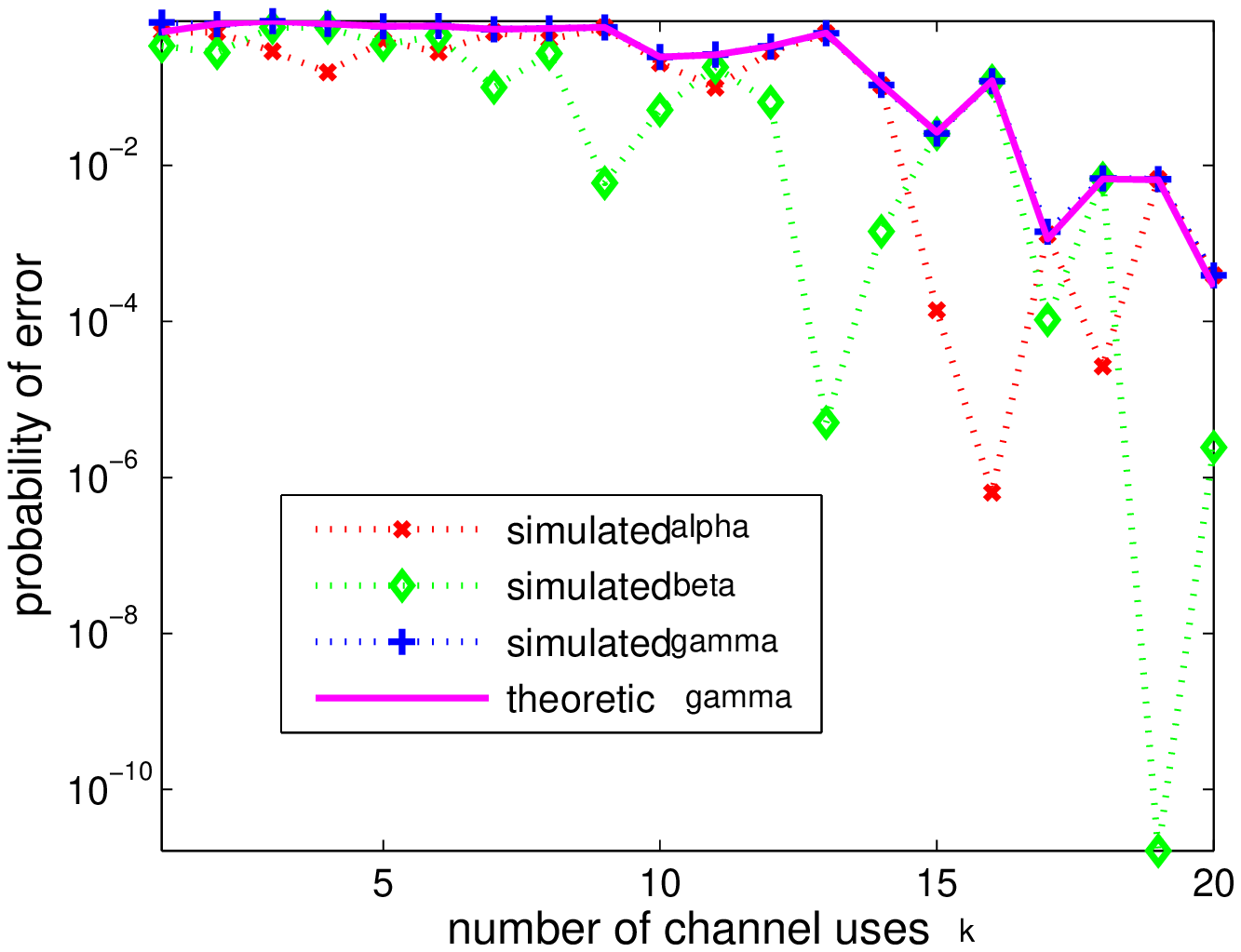}} }
%
 \psfrag{k}{\LARGE \bf $(K+1)$}
\center
 \subfigure[ ]{{\scalebox{.42}{\includegraphics{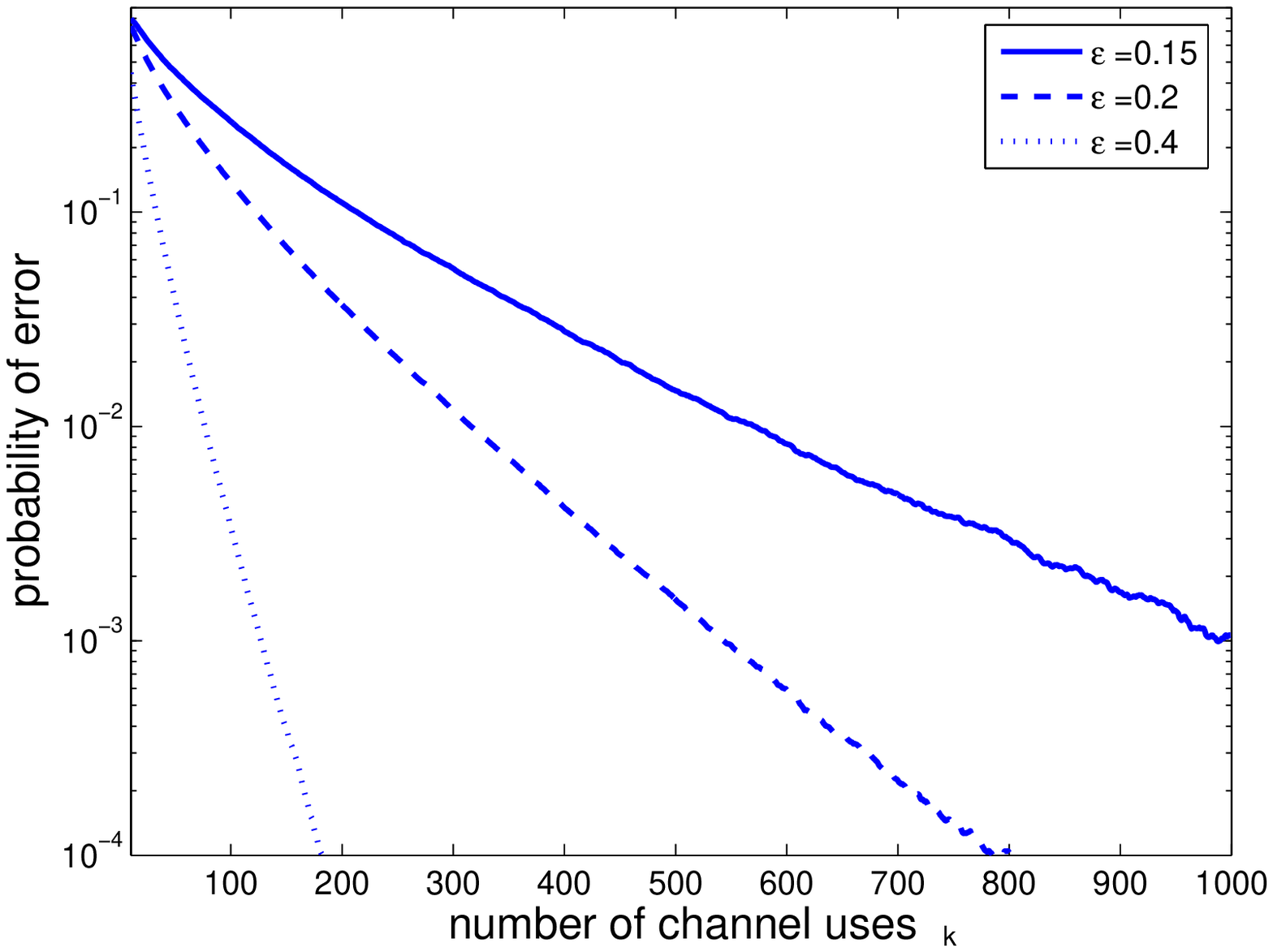}}}}
\hspace{0pt}
\psfrag{k }{\LARGE \bf $(K+1)$} \psfrag{kC}{\LARGE \bf $(K+1)C$}
\hspace{-0pt} \subfigure[
]{{\scalebox{.38}{\includegraphics{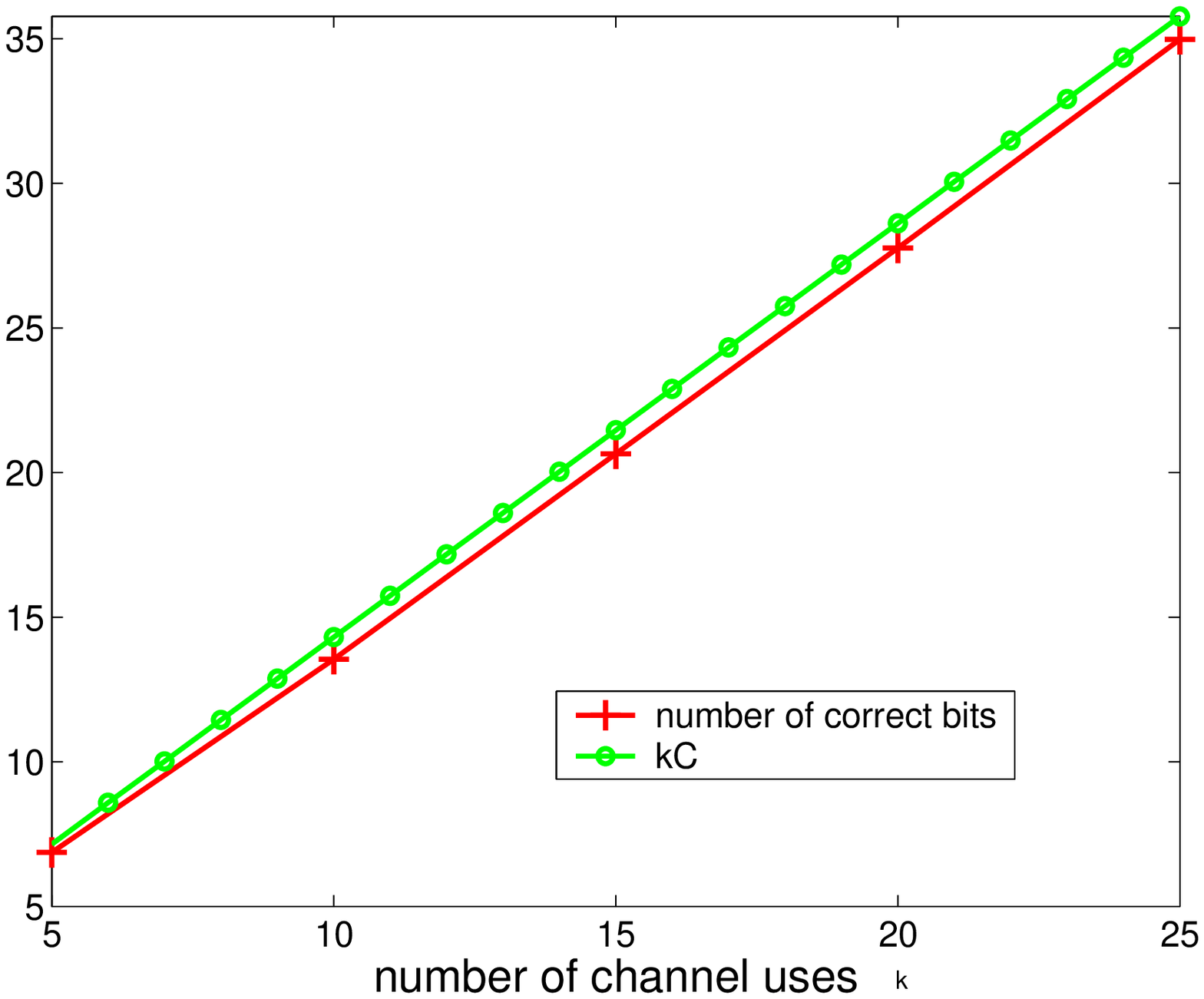}}}}
\caption{(a) The channel state transitions of an Gilbert-Elliot fading channel.
(b) The simulated $PE_{K|S}^{(j)}$, simulated $PE_{K|S}$, and theoretic
$PE_{K|S}$.
(c) The theoretic ${PE}_K$.  (d) The number of bits that has
been correctly decided and the number of bits that could be
correctly decided if at each step the capacity rate is attained.  It is assumed that
$s[1]=2$, $s[2]=1$, $p_{11}=0.65$, $p_{22}=0.38$, $P=3$, and
$\ep=0.2$ (i.e. $R=0.8C$), unless otherwise specified in the legend.
}\label{fig:simu}
\end{figure}

\section{Conclusions and Future Work} \label{sec:conclude}

In this paper, we proposed capacity-achieving feedback
communication schemes for, first, an FSMC with CSI available to the transmitter without delay (i.e. $\calF_{TCSI}$), second, an i.i.d. infinite-state fading channel with CSI available to the transmitter with a unit delay (i.e. $\calF_{II,DTCSI}$), and third, an FSMC with CSI available to the transmitter with a unit delay (i.e. $\calF_{DTCSI}$).  Instantaneous receiver-side CSI is always assumed for all the channels.
We established the
equivalence between feedback stabilization over a time-varying fading channel and
communication with access to noiseless output feedback over the same
channel.  We have shown that the control-oriented perspective may be used to facilitate the
study of feedback communication.

There are several open directions for future work.  First, we wish to relax the
assumption of perfect CSI and obtain the optimal strategies for channels with perfect output feedback and imperfect CSI. 
The assumption of perfect output feedback also needs to be relaxed but we remark that this is a longstanding challenge (cf. e.g. \cite{kailath1,kailath2}).
In addition, we wish to further explore the role of the cheap control (or
its counterpart in estimation theory, the Kalman filter) in feedback
communication, which may reveal further tight connections among communication, estimation, and control (cf. \cite{liu:phd}).
\appendix

\subsection{On different system formulations} \label{app:variations}

We comment on the relationship and differences between our formulation of the communication scheme for the AWGN channel (see Sec. \ref{sub:awgnscheme}) and other popular SK-type feedback communication schemes in their original forms.  These comments also apply to the proposed schemes for the more general channel $\calF$.  
First, our formulation is essentially the scheme studied in \cite{gallager}; the only difference is whether an extra operation is used to remove the estimation bias (see p. 481 \cite{gallager} and Remark \ref{rem:unbiased}).
Second, our formulation does not
involve unbounded coding parameters or unbounded signal power (from (\ref{eq:states1}) one can see that all the moments of the system state $x_k$ is bounded, and hence other signals, e.g., $u_k$ and $y_k$, which are linearly dependent on $x_k$, are also bounded), whereas the one in
\cite{kailath1} involves exponentially growing bandwidth,
the one in \cite{kailath2} involves an exponentially growing parameter
$\alpha^k$ where $\alpha>1$ and $k$ denotes the time index, and the one in \cite{elia_c5} generates a feedback signal with exponentially
growing power, despite the facts that they all generate the same channel inputs, same outputs, and same decoded messages, and that one formulation can be obtained as a simple reformulation of others. Third, our formulation
differs from the original SK scheme in that, ours
performs the same operation at every step, whereas the original SK formulation
performs its startup operation different from later steps. Although
ours has the advantage of unifying the operations for all
steps (which simplifies the control-oriented analysis), it has to either remove the bias term using an extra
operation (\cite{gallager} and Remark \ref{rem:unbiased}) or wait long enough until that exponentially
vanishing bias becomes negligible (Section IV of
\cite{elia_c5}). In contrast, the original SK scheme is unbiased since
the special startup operation eliminates the bias.

\subsection{The converse proof} \label{app:converse}

This proof is motivated by the converse proofs in \cite{vis99} and \cite{periodfb07}.

For any $(M_K,K+1)$ code with the message $W_K$ uniformly randomly selected from the set $w_{K}$, the Fano's inequality yields that
\begin{equation} \ba{lll} h(PE_K)+PE_K \log M_K &\geq& h(W_K | y_0^K, S_0^K ) \\ & = & h(W_K) - I(W_K;y_0^K, S_0^K ) \\  & = & \log M_K - I(W_K;y_0^K, S_0^K )   \ea
\end{equation}
and hence that
\begin{equation}  R_K:=\frac{1}{K+1}\log M_K \leq \frac{1}{(K+1)(1-PE_K)} \left(h(PE_K) + I(W_K;y_0^K, S_0^K ) \right). \end{equation}
If a sequence of $(M_K,K+1)$ codes leads to that $PE_K \rightarrow 0$ and hence $h(PE_K) \rightarrow 0$, then the sequence of rates $R_K$ must satisfy
\begin{equation} \liminf _{K \rightarrow \infty} R_K \leq \limsup _{K \rightarrow \infty} \frac{1}{K+1} I(W_K;y_0^K, S_0^K ),  \end{equation}
which, by Definition \ref{def:achievable_rate}, implies that, for any achievable rate $R$,
\begin{equation} R \leq \limsup _{K \rightarrow \infty} \frac{1}{K+1} I(W_K;y_0^K, S_0^K ).  \end{equation}
In addition, each $(M_K,K+1)$ code must satisfy the power constraint
\begin{equation}  \frac{1}{K+1} \E \sum_{k=0}^K P_{k}(W_K,y_0^{k-1},S_0^{k-d}) \leq \calP,
\label{pc:code} \end{equation}
in which for convenience we have defined $P_{k}(W_K,y_0^{k-1},S_0^{k-d}):=(u_k(W_K,y_0^{k-1},S_0^{k-d}))^2$.

It holds that
\begin{equation} \ba{lll}
& & I(W_K;y_0^K, S_0^K ) \\
&\eqa& I(W_K;y_0^K | S_0^K ) + I(W_K; S_0^K ) \\
& \eqb & I(W_K;y_0^K | S_0^K )  \\
&\eqc& h(y_0^K|S_0^K ) - h(y_0^K | S_0^K, W_K ) \\
&\eqd& \disp \sum _{k=0}^K  \left( h(y_k|S_0^{K},y_0^{k-1} ) - h(y_k|S_0^{K},y_0^{k-1},W_K) \right) \\
&=& \disp \sum _{k=1}^K  \left( h(y_k|S_0^{K},y_0^{k-1} ) - h(y_k|S_0^K,y_0^{k-1},W_K ) \right) + h(y_0|S_0 ) - h(y_0|S_0,W_K) ,
\ea  \label{eq:converse1} \end{equation}
where (a) is due to the chain rule of mutual information, (b) follows from the independence between $W_K$ and $S_0^K$, and (c) and (d) follows from definitions.

Note that for the first term in the last line of (\ref{eq:converse1}) we have that
\begin{equation} \ba{lll}
h(y_k|S_0^{K},y_0^{k-1}) &\ineqa&  h(y_k|S_0^{k-d},S_k,y_0^{k-1}) \\
&\eqb& h\left(S_k u_k(W_K,S_0^{k-d},y_0^{k-1})+N_k|S_0^{k-d},S_k,y_0^{k-1} \right) \\
&\ineqc& \disp \frac{1}{2} \E \log  2\pi e \E \left( S_k u_k(W_K,S_0^{k-d},y_0^{k-1})+N_k |S_0^{k-d},y_0^{k-1},S_k\right) ^2 \\
&=& \disp  \frac{1}{2} \E \log  2\pi e  \left((S_k)^2 \E(u_k(W_K,S_0^{k-d},y_0^{k-1})|S_0^{k-d},y_0^{k-1})^2 + 1 \right) \\
&=& \disp  \frac{1}{2} \E \log  2\pi e  \left((S_k)^2  \E P_{k}(W_K,y_0^{k-1},S_0^{k-d}|S_0^{k-d},y_0^{k-1}) + 1 \right)
,
\ea \end{equation}
where (a) is because conditioning reduces entropy, (b) is due to the definition of $y_k$, and (c) is because Gaussian distribution maximizes entropy (with equality if $u_k(W_K,S_0^{k-d},y_0^{k-1})$ given $(S_0^{k-d},y_0^{k-1})$ is Gaussian.  For the second term in the last line of (\ref{eq:converse1}) we have that
\begin{equation} \ba{lll} h(y_k|S_0^K,y_0^{k-1},W_K ) &=& h(y_k|u_k,S_k,S_0^K,y_0^{k-1},W_K ) \\
&=& \disp h(N_k ) =\frac{1}{2}\log 2\pi e .  \ea \end{equation}

Therefore, we obtain that
\begin{equation} \ba{lll}
& & I(W_K;y_0^K, S_0^K ) \\
&\leq& \disp \sum _{k=1}^K  \frac{1}{2} \E  \log  \left(1+(S_k)^2 \E P_{k}(W_K,y_0^{k-1},S_0^{k-d}|S_0^{k-d},y_0^{k-1}) \right)  + I(W_K;y_0|S_0) \\
&\eqa& \disp \sum _{k=1}^K  \frac{1}{2} \E \left. \left\{ \E \left[ \log  \left(1+(S_k)^2  \E P_{k}(W_K,y_0^{k-1},S_0^{k-d}|S_0^{k-d},y_0^{k-1}) \right) \right| S_{k-d},S_k \right] \right\}  + I(W_K;y_0|S_0) \\
&\ineqb& \disp \sum _{k=1}^K  \frac{1}{2} \E  \log  \left[1+(S_k)^2  \E \left. \left(\E P_{k}(W_K,y_0^{k-1},S_0^{k-d}|S_0^{k-d},y_0^{k-1}) \right|S_{k-d},S_k \right)  \right]  + I(W_K;y_0|S_0) \\
&\eqc& \disp \sum _{k=1}^K  \frac{1}{2} \E  \log  \left[1+(S_k)^2  \E P_{k}\left(W_K,y_0^{k-1},S_0^{k-d}|S_{k-d} \right)  \right]  + I(W_K;y_0|S_0) \\
&:=& \disp \sum _{k=1}^K  \frac{1}{2} \E  \log  \left(1+(S_k)^2  \gamma(S_{k-d} )  \right)  + I(W_K;y_0|S_0) ,
\ea \end{equation}
in which (a) is due to the law of total expectation, (b) follows from Jensen's inequality, and (c) is because of the law of total expectation and the Markov property that $S_0^{k-d-1}$ and $y_0^{k-1}$ are independent of $S_k$ if conditioned on $S_{k-d}$ when $d=1$ (when $d=0$ equality (c) obviously holds).

Thus, it holds that
\begin{equation}
R \leq  \limsup _{K \rightarrow \infty} \frac{1}{K+1} \left[\sum _{k=1}^K  \frac{1}{2} \E  \log  \left(1+(S_k)^2  \gamma(S_{k-d} )  \right)  + I(W_K;y_0|S_0) \right],
\end{equation}
subject to power constraint $\sum_{k=0}^K \E \gamma(S_{k-d} ) \leq (K+1) \calP$.  By the stationarity and ergodicity of the channel state process, it holds that
\begin{equation} R \leq \frac{1}{2}\E    \log \left( 1 + (S_k)^2 \gamma(S_{k-d}) \right)    \end{equation}
where $S_{k-d}$ follows the stationary distribution and $\E \gamma(S_{k-d} ) \leq \calP$.  Finally we have $R \leq C$ by the optimality of $\Gamma(\cdot)$.

\subsection{Proof for the channel $\calF_{II,DTCSI}$} \label{app:inf}

One can easily compute that the asymptotic signaling rate is $R=(1-\epsilon) \log \tilde{a}$.

To verify the power constraint, assume a fixed channel state sequence $S_0^k$, and in particular, $S_{k-1}=s[l]$. Then from the closed-loop dynamics one can derive that
\be  \E (x_{k}|S_0^k)^2- \calP = a(s[l])^{-2} \left(\E (x_{k-1}|S_0^{k-1})^2-\calP \right) . \ee
%
%
Since $(x_0)^2 \leq \calP$ and $a(s[l]) \geq 1$, by induction $\E (x_{k}|S_0^k)^2 \leq \calP$ for any $k$.  Hence for any $k$, $\E (x_{k})^2$ and the average input power is no larger than $\calP$.

As $S_t$ forms an i.i.d. process, so does $\log a(S_t)$ and thus it holds that
\begin{equation}  \frac{1 } {K+1} \sum_{k=0}^K \log a(S_k) \arrowP \log \tilde{a}>0 .
\label{aep:iid} \end{equation}

From the end-to-end equation (\ref{eq:hbfxiid}), we have
\begin{equation} \hatx_{0,K|S,x_0} \sim \mathcal{N}\left( \left(1 - (\phi_K)^2 \right) x_0, \left( \phi_K \psi_K  \right) ^2 \right) , \label{dist:iid} \end{equation}
in which
\be \ba{lll} (\psi_K)^2 &:=& \disp \sum_{t=0}^K ( \phi_K)^2 (\phi_t)^{-2} b(S_{t})^2  \\
&\ineqa& \disp \sum_{t=0}^K  b(S_{t})^2 \\
&=& \disp \sum_{t=0}^K \calP (1-a(S_t)^{-2}) \\
&\leq& (K+1) \calP  \ea \label{ineq:psi}\ee
where (a) is because $\phi_K \phi_t = \prod_{i=t+1}^K a(S_{i})^{-1} \leq 1$. 

Similar to the case for the channel $\calF_{TCSI}$, to prove that the probability of error $PE_K$ decays to zero, it is sufficient to show that $PE_{K|S} \arrowP 0$.
From (\ref{dist:iid}) and (\ref{ineq:psi}), it holds that
\begin{equation}  PE_{K|S,x_0}
\leq \disp 2 Q\left(   \frac{ 1}{  \sqrt{K+1} \phi_K \left(2 \exp ( (K+1)(1-\ep) \log \tilde{a})  - 1 \right)}  -  \frac{\phi_K^{(j)} }
{ \sqrt{K+1}}   \right) .
\end{equation}
By (\ref{aep:iid}), we can show $\phi_K \arrowP 0$ since $(\log \phi_K)/(K+1) \arrowP -\log \tilde{a}$.  It then suffices to show
\be \eta_K:=\phi_K  \exp \left( (K+1)(1-\ep) \log \tilde{a} \right) \arrowP 0 \ee
in order to prove $PE_{K|S,x_0} \arrowP 0$ (note that $\phi_K$ decays faster than $\eta_K$).  However
\be \ba{lll} \disp \frac{1}{K+1} \log\eta_K &=& \disp  -\frac{1 } {K+1} \sum_{k=0}^K \log a(S_k)  + (1-\ep)\log \tilde{a}   \\
\disp &\arrowP&  \disp  -\ep \log \tilde{a}  <0 . \ea  \ee
Therefore, we conclude that $PE_K \rightarrow 0$.

\subsection{Proof for the channel $\calF_{DTCSI}$} \label{sub:rate}

It is straightforward to compute $R=(1-\epsilon) \log \tilde{a}$.

To verify the power constraint, similar to the case for the channel $\calF_{TCSI}$, one can show that the recursion for $\E (x_k^{(j)})^2$ satisfies
\beas  \E (x_{k}^{(j)}|S_0^k)^2- \Gamma(s[j])
&=& a(s[j],s[l])^{-2} \left(\E (x_{k-1}^{(j)}|S_0^k)^2-\Gamma(s[j])\right) , \eeas
and thus $\E (x_{k}^{(j)}|S_0^k)^2 \leq \Gamma(s[j])$ for any $k$ and any $j$.  Over all possible channel realizations, $\Sigma_i$ is active with probability $\pi[i]$.  Since $\sum \pi[i] \Gamma[i] \leq \calP$, the average input power constraint is satisfied.

%
To show the vanishing probability of error $PE_K$, as proven before it is sufficient to show that $PE_{K|S}^{(j)} \arrowP 0$ for all $j$. From the end-to-end equation (\ref{eq:hbfx}) it holds that
%
%
for each $j$, 
\begin{equation} \hat{x}_{0,K|S,\bfx_0}^{(j)} \sim \mathcal{N} \left( (1 -  (\phi_K^{(j)})^2) x_{0}^{(j)} , \left(  \phi_K^{(j)} \psi_K^{(j)}   \right)^2
 \right) \label{distn:xj}\end{equation}
where
\be  \ba{lll} \left(\psi_K^{(j)} \right) ^2 &:=& \disp \sum_{S_{k-1}=s[j],k \in \{0,...,K\}} (\phi_K)^{2} (\phi_k)^{-2}  b(s[j],S_{k})^2  \\
&\leq& \disp \sum_{S_{k-1}=s[j],k \in \{0,...,K\}}   b(s[j],S_{k})^2  \\
&=& \disp \sum_{S_{k-1}=s[j],k \in \{0,...,K\}}  \Gamma(s[j]) (1- a(s[j],S_{k})^{-2})  \\
&\leq&  (K+1)  \Gamma(s[j])  . \ea
 \ee
If $\Gamma(s[j])=0$ (which is equivalent to $\bar{a}[j]=1$ according to Lemma \ref{lemma:bara}), as shown in the TCSI case, we have $PE_{K|S}^{(j)}=0$.  So we focus on the case with $\Gamma(s[j])>0$, i.e. $\bar{a}[j]>1$. It holds that
\begin{equation} \ba{lll} PE_{K|S,\bfx_0}^{(j)} 
&\leq& \disp 2Q\left(   \frac{ 1}{   \sqrt{K+1}\phi_K^{(j)} \left(2\bara[j]^{(K+1)(1-\epsilon)} - 1 \right)}  -  \frac{\phi_K^{(j)} }
{\sqrt{K+1}}   \right)   .
\ea
\end{equation}
However, since
\begin{equation}  \ba{lll}
 \disp \frac{1}{K+1} \log \phi_K^{(j)} &=& \disp  \sum_{l=1}^m \frac{-n(j,l,K)}{K+1}\log a(s[j],s[l] )\\
& \arrowP & \disp \sum_{l=1}^m -\pi[j] p_{jl} \log a(s[j],s[l] )  \\
&=& -\log \bara[j] < 0  ,\ea \end{equation}
it holds that $\phi_K^{(j)} \arrowP 0$.
%
In addition, letting $\eta_K:=\sqrt{K+1}\bara[j]^{(K+1)(1-\epsilon)} \phi_K^{(j)}$, we can show that
\begin{equation} \ba{lll}  \disp  \frac{1}{K+1} \log \eta_K
&=& \disp    \frac{1}{2(K+1)} \log (K+1) + \sum _{l=1}^m \left( (1-\epsilon)\pi[j] p_{jl} - \frac{n(j,l,K)}{K+1} \right) \log a(s[j],s[l] )      \\
&\arrowP& \disp     \sum _{l=1}^m -\epsilon \pi[j] p_{jl}  \log a(s[j],s[l] )      \\
&=& \disp   - \ep \log \bar{a}[j] < 0
\ea  \label{Q:den}
\end{equation}
and hence $\eta_K  \arrowP 0$.  Clearly $\sqrt{K+1}\phi_K^{(j)} \arrowP 0$ and it decays faster than $\eta_K$ does. Thus we conclude that $PE_{K|S,\bfx_0}^{(j)}  \arrowP 0$, $PE_{K|S}^{(j)} \arrowP 0$, and $PE_k \rightarrow 0$.

\vspace{.1in} \hspace*{2.5in} {{\textbf{\noindent \small{ACKNOWLEDGMENT}}}}

The authors wish to thank Zhengdao Wang, Anant Sahai, Krishna
Athreya, Amos Lapidoth, the associated editor, and anonymous
reviewers for useful discussions and suggestions.

\hfill \markright{\textsf{References}} \small

\end{document}